\documentclass[aps,preprint,showpacs]{revtex4-1}
\usepackage{amsfonts}
\usepackage{amsmath}
\usepackage{amssymb}
\usepackage{graphicx}
\usepackage{rotating}
\usepackage{hyperref}

\begin{document}

\title{Hadronic and hybrid stars subject to density dependent magnetic fields}
\author{R. Casali}\email[]{rcasali@fisica.ufsc.br}
\author{L. B. Castro}\email[ ]{luis.castro@pgfsc.ufsc.br}
\author{D. P. Menezes}\email[ ]{debora.p.m@fsc.ufsc.br}
\affiliation{Departamento de F\'{\i}sica CFM,
Universidade Federal de Santa Catarina, Florian\'{o}polis, SC CP. 476, CEP 88.040-900, Brazil}


\begin{abstract}
In the light of the very massive neutron stars recently detected and the new
possible constraints for the radii of these compact objects, we
revisit some equations of state obtained for hadronic and hybrid stars under
the influence of strong magnetic fields.
We present our results for hadronic matter taking into account the effects
of the inclusion of anomalous magnetic moment.
Additionally, the case of hybrid stars under the influence of strong magnetic fields is considered. We study the structure of hybrid stars based on the Maxwell condition (without a mixed phase), where the hadron phase is described by the non-linear Walecka model (NLW) and the quark phase by the Nambu-Jona-Lasinio model (NJL). The mass-radius relation for each case are calculated and discussed. We show that both hadronic and hybrid stars can bear very high masses and radii
compatible with the recently observed high mass neutron stars.
\end{abstract}

\maketitle

\section{Introduction}

The study of neutron stars provides an excellent laboratory for the understanding of dense matter under extreme conditions. A typical neutron star has a mass of the order of $1-2~M_{\odot}$ and a radius of the order of $11$~Km, its
temperature stands around $11^{11}$~K right after its birth, followed by a
rapid cooling process led by neutrino emission.
Although the conventional models of neutron stars assume that dense matter is composed of hadrons and leptons, as the density increases inwards, the Fermi level of the nucleons increases to values above the mass threshold of heavier particles, opening the possibility that another particle is created, reducing the total energy. Baryon number conservation, violation of strangeness and the Pauli exclusion principle guarantee this mechanism. The same phenomenon is responsible for the reduction of the total pressure.
On the other hand, the Bodmer-Witten conjecture \cite{Bodmer,Witten, Itoh} states that quarks could be deconfined from the hadrons, forming a stable quark matter. This raises questions about the true constitution of ground state matter at high densities \cite{collins, glen} and arises the possibility that compact stars can be constituted of pure deconfined quark matter or perhaps be hybrid stars, containing in their core a pure quark phase or a non-homogeneous mixed quark-hadron phase, whose existence is a source of intense discussions in the literature \cite{glen,Stellar-matter-with,mene3,bielich,pagli,lugo2,scher,baldo,shov,buba,kahn,steiner,yang}. Most neutron stars have masses of the order
of $1.4 M_\odot$, but more recently, at least two pulsars, PSR J1614-2230 \cite{demorest} and PSR J0348+0432 \cite{antoniadis} were confirmed
to bear masses of the order of $2 M_\odot$. If one believes that a unique equation of state (EOS) has to be able to account for all possible observational data, a hard EOS at high densities is then mandatory.

Neutron stars generally manifest themselves as pulsars, which are powered by their rotation energy or as accreting X-ray binaries, which are
powered by the gravitational energy. Some compact objects, known as magnetars, do not fit into any of these categories. They are normally
isolated neutron stars whose main power source is the magnetic field and two classes have been discovered: the soft gamma-repeaters that are x-ray transient sources and the anomalous x-ray pulsars, a class of persistent x-ray sources with no sign of a binary companion. Although not very many magnetars have been confirmed so far (around two dozens), they are expected to constitute up to 10\%
of the total neutron star population.
Hence, magnetars are extremely magnetized neutron stars, with magnetic fields reaching $B=10^{15}~G$ at the surface and central magnetic fields that could reach even higher values \cite{Kouveliotou, Duncan}.

At such high range of magnitudes magnetic fields can interfere on the thermodynamic and hydrodynamic properties \cite{Chaichian,Martinez2,Paulucci,Martinez}, causing anisotropy. According to calculations with free fermion systems at
zero \cite{vivian_2010} and finite temperature \cite{nosso_2012}, an upper
limit for the value of the magnetic field can be established if anisotropic
effects are to be disregarded. This upper limit depends both on the
temperature of the system and on the inclusion (or not) of anomalous magnetic
moments. Temperature washes out anisotropic effects on the pressure of the
system and anomalous magnetic moments enhance them \cite{nosso_2012}.
If the same problem is tackled in a system subject to stellar matter conditions,
one has to take into account not only anisotropic effects due to matter
contribution \cite{guang}, but also due to the pure magnetic field
contribution that arises from the eletromagnetic tensor
\cite{Paulucci,nosso_2013,sinha}.
In this case, calculations done with different models indicate that the
maximum magnetic field that can be used in order to avoid
anisotropic effect is
of the order of $B=10^{18}$ G, for a baryonic chemical potential equal to
$\mu_B=1500$ MeV in zero temperature systems. Nevertheless, when a magnetic
field that varies with density (or analogously, with the baryonic chemical
potential) is considered as in \cite{bandy}, a slightly higher magnetic
field can be considered
because stability is maintained up to higher densities \cite{nosso_2013}.
Hence, in the present work we consider the maximum possible magnetic field that
can still be used so that anisotropic effects can be neglected
and it is of the order of $B=10^{18.5}$ G $\simeq 3.1 \times 10^{18}$ G. One
should bear in mind that these very high densities (corresponding to
a baryon chemical potential of the order of 1500 MeV) are not always reached
in the core of hadronic and hybrid stars.
An important consideration in the choice of the values of reasonable magnetic
fields is the error in the use of the Tolman-Oppenheimer-Volkoff equations,
valid only for homogeneous and isotropic systems. According to a recent estimate
\cite{mallick}, the error in the calculation of the stellar mass is very small,
i.e., around $10^{-4}-10^{-3} M_\odot$ for fields of the order of $B=10^{18}$ G and
$10^{-2}-10^{-1} M_\odot$ for fields of the order of $B=10^{19}$ G. Hence, we
believe that our choice for the maximum magnetic field as
$3.1 \times 10^{18}$ G is very reasonable.

Another possible complication that the introduction of a magnetic field can bring is that it is a source of gravitational energy. The virial theorem sets an upper limit so that magnetic fields do not imply on a gravitational collapse of the magnetar.  As said above, we apply a density dependent magnetic field
\cite{bandy} on our equations, and this ensures no gravitational setbacks.

The influence of strong magnetic fields on the quark-hadron phase transition was first discussed in \cite{bandy}, using a Dirac-Hartree-Fock approach within a mean-field approximation to describe both the hadronic and the quark phases. For the hadronic matter a system of protons, neutrons and electrons was considered, and for the quark phase the MIT bag model with one-gluon exchange was used. A very hard quark equation of state (EOS) was obtained so that the hybrid star did not have a quark core. The authors concluded that compact stars have a smaller maximum mass in the presence of strong magnetic fields, a result that does not agree with other more recent works where hadronic stars \cite{cardall,bro,lopes,jame2} or quark stars \cite{mene2,mene} in the presence of strong magnetic fields have been studied.

If strong magnetic fields are considered, contributions from the anomalous magnetic moments (AMM) of the nucleons and hyperons should also be taken into account. Experimental measurements find that  $k_{p}=\mu_{N}(g_{p}/2-1)$ for protons and $k_{n}=\mu_{N}g_{p}/2$ for neutrons, where $\mu_{N}$ is the nuclear magneton, $g_{p}=5.58$ and $g_{n}=-3.83$ are the Landau g-factors of protons and neutrons, respectively. In \cite{bro2} the proton and neutron AMM controbutions to hadronic EOS were computed for the first time and later, it was extended to include the contribution from the eight lightes baryons \cite{bro}. Although the general conclusion was that a strong magnetic field softens the EOS, which is not true if
the pure electromagetic field contribution is adequately included, the authors
pointed out that the AMM stiffens the EOS.  This problem was then revisited and
the EOS was obtained with a density dependent model and the inclusion of the
scalar-isovector $\delta$ mesons, which were shown to be important for low
mass stars \cite{Stellar-matter-with}. In neither of these works the
stellar maximum masses were computed and the magnetic fields considered were
always very high.

In the present work we first study magnetars composed of hadronic matter only. We consider the inclusion of the anomalous magnetic moments of all the particles in the baryonic octet and its effects on stellar properties. We describe the hadronic matter within the framework of the relativistic non-linear Walecka model (NLW) \cite{Serot}. We also consider a magnetic field that increases, in a density dependent way, from the surface ($10^{15}~G$) to the interior of the star.
Comparing our work with \cite{lopes} we show that  significant differences on maximum masses and their respective radii for stronger magnetic fields can be obtained depending on the choice of the parameters for slow and fast decays of the density dependent magnetic field.
This means that different combinations of parameters can generate controllable values for masses and radii, as expected from the results obtained in
\cite{sinha}.

Additionally we study hybrid stars under the influence of magnetic fields. The structure of hybrid stars is based on the Maxwell condition (without mixed phase), the hadronic matter is again described by the NLW \cite{Serot} and the quark matter by the Nambu-Jona-Lasinio (NJL) model \cite{njl} composed of quarks up ($u$), down ($d$) and strange ($s$) in $\beta$-equilibrium. We also assume the density-dependent magnetic field and we choose the same two sets of values for the parameters $\beta$ and $\gamma$ for slow and fast decays of the density dependent magnetic field as in the hadronic case. We show that hybrid stars have a larger maximum mass in the presence of strong magnetic field as compared with the
results presented in \cite{marce} and that the slow decays produce smaller maximum masses, but larger radii. Our result for the values of maximum masses and radii for a weak magnetic field are in agreement with the results obtained for $B=0~G$ \cite{marce}. The macroscopic properties of hybrid stars under effects of magnetic fields have already been studied in the literature \cite{panda}, where the quark phase was described by the MIT bag model. A qualitative analysis shows that the properties of the stars obtained in both cases are
very similar, but a quantitative analysis shows that we obtain higher maximum masses and radii for weaker magnetic fields with the NJL model describing the quark core. This consideration is model and parameter dependent and, hence, has
to be taken with care. We discuss these differences again when we present our
results.

The organization of this work follows: in Sec. II, we give a brief review of the formalism used to describe the hadronic and quark phases under a magnetic fieldand discuss the conditions for building of a hybrid star with the Maxwell
construction. In Sec. III we present our results for the inclusion of a density dependent magnetic field on the total energy density and total pressure, particle fractions and mass-radius relation for hadronic and hybrid stars. Finally, in Sec. IV we present our main conclusions.

\section{Formalism}

In this section we present an overview of the equations used to describe the hadronic (subsection A), quark (subsection B) and hybrid (subsection C)  phases. We describe hadronic matter within the framework of the relativistic non-linear Walecka model (NLW) \cite{Serot}. The quark matter is described by SU(3) version of the the Nambu-Jona-Lasinio (NJL) model \cite{Buballa}. Hybrid matter is
built using the Maxwell conditions.

\subsection{Hadronic phase under a magnetic field}
\label{sec:A}

For the description of the equation of state (EOS) of hadronic matter, we employ a field-theoretical approach in which the baryons interact via the exchange of $\sigma-\omega-\rho$ mesons in the presence of a magnetic field $B$ along the $z-$axis. The total lagrangian density reads:

\begin{equation}\label{lt}
    \mathcal{L}_{H}=\sum_{b}\mathcal{L}_{b}+\mathcal{L}_{m}+\sum_{l}\mathcal{L}_{l}+\mathcal{L}_B\,.
\end{equation}

\noindent where $\mathcal{L}_{b}$, $\mathcal{L}_{m}$, $\mathcal{L}_{l}$ and $\mathcal{L}_{B}$ are the baryons, mesons, leptons and electromagnetic field Lagrangians, respectively, and are given by
\begin{eqnarray}\label{lb}
    \mathcal{L}_{b} &=& \overline{\psi}_{b}\left(i\gamma_{\mu}\partial^{\mu}-q_{b}\gamma_{\mu}A^{\mu}-m_{b}+g_{\sigma b}\sigma \right. \nonumber \\
    && \left. -g_{\omega b}\gamma_{\mu}\omega^{\mu}-g_{\rho b}\tau_{3b}\gamma_{\mu}\rho^{\mu}-k_{b}\sigma_{\mu\nu}F^{\mu\nu}\right)\psi_{b}\,,
\end{eqnarray}
\begin{eqnarray}\label{lm}
   \mathcal{L}_{m} &=& \frac{1}{2}(\partial_{\mu}\sigma\partial^{\mu}\sigma-m_{\sigma}^{2}\sigma^{2})-U(\sigma)+
    \frac{1}{2}m_{\omega}^{2}\omega_{\mu}\omega^{\mu} \nonumber \\
   && -\frac{1}{4}\Omega_{\mu\nu}\Omega^{\mu\nu}+
    \frac{1}{2}m_{\rho}^{2}\vec{\rho}_{\mu}\cdot\vec{\rho}_{\mu}-\frac{1}{4}P^{\mu\nu}P_{\mu\nu} \,,
\end{eqnarray}
\begin{equation}\label{ll}
    \mathcal{L}_{l}=\overline{\psi}_{l}\left(i\gamma_{\mu}\partial^{\mu}-q_{l}\gamma_{\mu}A^{\mu}-m_{l}\right)\psi_{l} \,,
\end{equation}
\begin{equation}\label{lcm}
    \mathcal{L}_{B}=-\frac{1}{4}F^{\mu\nu}F_{\mu\nu} \,.
\end{equation}

\noindent where he $b$-sum runs over the baryonic octet $b\equiv N~(p,~n),~\Lambda,~\Sigma^{\pm,0},~\Xi^{-,0}$, $\psi_{b}$ is the corresponding baryon Dirac field, whose interactions are mediated by the $\sigma$ scalar, $\omega_{\mu}$ isoscalar-vector and $\rho_{\mu}$ isovector-vector meson fields. The baryon mass and isospin projection are denoted by $m_{b}$ and $\tau_{3b}$, respectively, and the masses of the mesons are $ m_{\sigma}= 512~$MeV, $m_{\omega}=783~$MeV and $m_{\rho}=770~$MeV. The strong interaction couplings of the nucleons with the meson fields are denoted by $g_{\sigma N}=8.910$, $g_{\omega N}=10.610$ and $g_{\rho N}=8.196$. We consider that  the couplings of the hyperons with the meson fields are fractions of those of the nucleons, defining $g_{iH}=X_{iH}g_{iN}$, where the values of $X_{iH}$ are chosen as $X_{\sigma H}=0.700$ and $X_{\omega H}=X_{\rho H}=0.783$ \cite{glen}.
 The term $U(\sigma)=\frac{1}{3}\,bm_{n}(g_{\sigma N}\sigma)^{3}-\frac{1}{4}\,c(g_{\sigma N}\sigma)^{4}$ denotes the scalar self-interactions \cite{Boguta,Glandening1,Glandening2}, with
 $c=-0.001070$ and $b=0.002947$. The mesonic and electromagnetic field tensors are given by their usual expressions $\Omega_{\mu\nu}=\partial_{\mu}\omega_{\nu}-\partial_{\nu}\omega_{\mu}$, ${\bf P}_{\mu\nu}=\partial_{\mu}\vec{\rho}_{\nu}-\partial_{\nu}\vec{\rho}_{\mu}-g_{\rho b}(\vec{\rho}_{\mu}\times\vec{\rho}_{\nu})$ and $F_{\mu\nu}=\partial_{\mu}A_{\nu}-\partial_{\nu}A_{\mu}$. The baryon anomalous magnetic moments (AMM) are introduced via the coupling of the baryons to the electromagnetic field tensor with $\sigma_{\mu\nu}=\frac{i}{2}[ \gamma_{\mu},\gamma_{\nu} ]$ and the strength $\kappa_{b}=(\mu_{b}/\mu_{N})-q_{b}(m_{p}/m_{b})$, where $q_{p}$ and $m_{p}$ are the charge and mass of the proton and $\mu_{b}$ and $m_{b}$ are the magnetic moment and masses of the baryons, whose values can be seen in TABLE \ref{table3}. The $l$-sum runs over the two lightest leptons $l\equiv e,\mu$ and $\psi_{l}$ is the lepton Dirac field.
The symmetric nuclear matter properties at saturation density adopted in this work are given by the GM1 parametrization \cite{GM1}, with compressibility $K=300$ (MeV), binding energy $B/A=-16.3$ (MeV), symmetry energy $a_{sym}=32.5$ (MeV), slope $L=94$ (MeV), saturation density $\rho_{0}= 0.153~(fm^{-3})$ and nucleon mass $m=938$ (MeV).

\begin{table}[h]
 \centering
\begin{tabular}{|c|c|c|c|c|c|c|c|c|} \hline
 Baryon&p&n&$\Lambda^{0}$&$\Sigma^{+}$&$\Sigma^{0}$&$\Sigma^{-}$&$\Xi^{0}$&$\Xi^{-}$\\  \hline
$M_{b}$ (MeV)&938&938&1116&1193&1193&1193&1318&1318\\  \hline
$q_{b}$&1&0&0&1&0&-1&0&-1\\  \hline
$\mu_{b}/\mu_{N}$&2.79&-1.91&-0.61&2.46&-1.61&-1.16&-1.25&-0.65\\  \hline
$k_{b}$&1.79&-1.91&-0.61&1.67&-1.61&-0.38&-1.25&0.06\\  \hline
\end{tabular}
 \caption{\label{table3} Baryon masses, charges, magnetic moments and anomalous magnetic moments. We have taken negative values for the $\Sigma^{0}$ meson, contrary to \cite{bro,panda}.}
\end{table}

The following equations present the scalar and vector densities for the charged and uncharged baryons \cite{panda}, respectively:

\begin{eqnarray}
&&\label{densities1}\rho_{b}^{s}=\frac{|q_{b}|Bm^{\ast}_{b}}{2\pi^{2}}\sum_{\nu}^{\nu_{\mathrm{max}}}\sum_{s}\frac{\bar{m}_{b}^{c}}{\sqrt{m_{b}^{\ast2}+2\nu |q_{b}|B}}\ln\bigg|\frac{k_{F,\nu,s}^{\,b}+E_{F}^{\,b}}{\bar{m}_{b}^{c}} \bigg|\,,\\
&&\label{densities2}\rho_{b}^{v}=\frac{|q_{b}|B}{2\pi^{2}}\sum_{\nu}^{\nu_{\mathrm{max}}}\sum_{s}k_{F,\nu,s}^{\,b}\,,\\
&&\label{densities3}\rho_{b}^{s}=\frac{m^{\ast}_{b}}{4\pi^{2}}\sum_{s}\bigg[E_{F}^{\,b}k_{F,s}^{\,b}-\bar{m}_{b}^{2}\ln\bigg|\frac{k_{F,s}^{\,b}+E_{F}^{\,b}}{\bar{m}_{b}}\bigg|\bigg]\,,\\
&&\label{densities4}\rho_{b}^{v}=\frac{1}{2\pi^{2}}\sum_{s}\bigg[\frac{1}{3}(k_{F,s}^{\,b})^{3}-\frac{1}{2}s\mu_{N}k_{b}B\bigg(\bar{m}_{b}\,k_{F,s}^{\,b}+(E_{F}^{\,b})^{2}\bigg(\arcsin\bigg(\frac{\bar{m}_{b}}{E_{F}^{\,b}}\bigg)-\frac{\pi}{2}\bigg) \bigg)\bigg]\,.
\end{eqnarray}

\noindent where $m_{b}^{\ast}=m_{b}-g_{\sigma}\sigma$, $\bar{m}_{b}^{c}=\sqrt{m_{b}^{\ast2}+2\nu |q_{b}|B}-s\mu_{N}k_{b}B$ and $\bar{m}_{b}=m_{b}^{\ast}-s\mu_{N}k_{b}B$.  $\nu=n+\frac{1}{2}-$sgn$(q_{b})\frac{s}{2}=0,1,2,...$ are the Landau levels for the fermions with electric charge $q_{b}$, $s$ is the spin and assumes values
$+1$ for spin up and $-1$ for spin down cases.

The energy spectra for the baryons are given by \cite{bro2,Stellar-matter-with}:
\begin{eqnarray}\label{energy}
&&E_{\nu,s}^{\,b}=\sqrt{(k_{z}^{\,b})^{2}+(\sqrt{m_{b}^{*2}+2\nu |q_{b}|B}-s\mu_{N}k_{b}B)^{2}}+g_{\omega b}\omega^{0}+\tau_{3b}g_{\rho b}\rho^{0}\\
&&E_{s}^{\,b}=\sqrt{(k_{z}^{\,b})^{2}+(\sqrt{m_{b}^{*2}+k_{\perp}^{2}}-s\mu_{N}k_{b}B)^{2}}+g_{\omega b}\omega^{0}+\tau_{3b}g_{\rho b}\rho^{0},
\end{eqnarray}
where $k_{\perp}=k_{x}+k_{y}$. The Fermi momenta $k_{F,\nu,s}^{\,b}$ of the charged baryons and $k_{F,s}^{\,b}$ of the uncharged baryons and their relationship with the Fermi energies of the charged baryons $E_{F,\nu,s}^{\,b}$ and uncharged baryons $E_{F,s}^{\,b}$ can be written as:
\begin{eqnarray}\label{Momentum}
&&(k_{F,\nu,s}^{\,b})^{2}=(E_{F,\nu,s}^{\,b})^{2}-(\bar{m}_{b}^{c})^{2} \\
&&(k_{F,s}^{\,b})^{2}=(E_{F,s}^{\,b})^{2}-\bar{m}_{b}^{2}.
\end{eqnarray}

For the leptons, the vector density is given by:
\begin{eqnarray}\label{vector_{density_{leptons}}}
&&\rho_{l}^{v}=\frac{|q_{l}|B}{2\pi^{2}}\sum_{\nu}^{\nu_{\mathrm{max}}}\sum_{s}k_{F,\nu,s}^{\,l},\\ \nonumber
\end{eqnarray}

\noindent where $k_{F,\nu,s}^{\,l}$ is the lepton Fermi momentum, which is related to the Fermi energy $E_{F,\nu,s}^{\,l}$ by:
\begin{eqnarray}\label{Momentuml}
&&(k_{F,\nu,s}^{\,l})^{2}=(E_{F,\nu,s}^{\,l})^{2}-\bar{m}_{l}^{2}\,, \qquad l=e,\mu,
\end{eqnarray}

\noindent with $\bar{m}_{l}=m_{l}^{2}+2\nu |q_{l}|B$.  The summation over the Landau level runs until $\nu_{\mathrm{max}}$, this is the largest value of $\nu$ for which the square of Fermi momenta of the particle is still positive and corresponds to the closest integer, from below to:
\begin{eqnarray}
&&\nu_{\mathrm{max}}=\bigg[\frac{(E_{F}^{\,l})^{2}-m_{l}^{2}}{2|q_{l}|B}\bigg], \qquad\mathrm{leptons}\label{ll1}\\
&&\nu_{\mathrm{max}}=\bigg[\frac{(E_{F}^{\,b}+s\mu_{N}k_{b}B)^{2}-m_{b}^{\ast2}}{2|q_{b}|B}\bigg], \qquad\mathrm{charged~baryons}.\label{ll2}
\end{eqnarray}

The chemical potentials of baryons and leptons are:
\begin{eqnarray}\label{chemicalp}
&&\mu_{b}=E_{F}^{\,b}+g_{\omega b}\omega^{0}+\tau_{3b}g_{\rho b}\rho^{0},\\
&&\mu_{l}=E_{F}^{\,l}=\sqrt{(k_{F,\nu,s}^{\,l})^{2}+m_{l}^{2}+2\nu |q_{l}|B}\,.
\end{eqnarray}

From the Lagrangian density~(\ref{lt}), and mean-field approximation, the energy density is given by
\begin{eqnarray}\label{energym}
\varepsilon_{m}= & & \sum_{b}(\varepsilon_{b}^c + \varepsilon_{b}^n)
+\frac{1}{2}m_{\sigma}\sigma_{0}^{2}\nonumber\\
&& +U(\sigma)+\frac{1}{2}m_{\omega}\omega_{0}^{2}+\frac{1}{2}m_{\rho}\rho_{0}^{2}\,,
\end{eqnarray}

\noindent where the expressions for the energy densities of charged baryons $\varepsilon_{b}^{c}$ and neutral baryons $\varepsilon_{b}^{n}$ are, respectively, given by:
\begin{eqnarray}\label{energy-densities-baryons}
\varepsilon_{b}^{c}& = &\frac{|q_{b}|B}{4\pi^{2}}\sum_{\nu}^{\nu_{\mathrm{max}}}\sum_{s}\bigg[k_{F,\nu,s}^{\,b}E_{F}^{\,b}+
(\bar{m}_{b}^{c})^{2}\ln\bigg|\frac{k_{F,\nu,s}^{\,b}+E_{F}^{\,b}}{\bar{m}_{b}^{c}}\bigg|\bigg],\label{ea1}\\
\varepsilon_{b}^{n}& = &\frac{1}{4\pi^{2}}\sum_{s}\bigg[\frac{1}{2}k_{F,\nu,s}^{\,b}(E_{F}^{\,b})^{3}
-\frac{2}{3}s\mu_{N}k_{b}B(E_{F}^{\,b})^{3}\bigg(\arcsin\left(\frac{\bar{m}_{b}}{E_{F}^{\,b}}\right)-\frac{\pi}{2}\bigg)\bigg.\nonumber\\
&&\bigg.-\bigg(\frac{1}{3}s\mu_{N}k_{b}B+\frac{1}{4}\bar{m}_{b}\bigg)\bigg(\bar{m}_{b}k_{F,\nu,s}^{\,b}E_{F}^{\,b}+
\bar{m}_{b}^{3}\ln\bigg|\frac{E_{F}^{\,b}+k_{F,\nu,s}^{\,b}}{\bar{m}_{b}}\bigg|\bigg)\bigg]\,.\label{ea2}
\end{eqnarray}

\noindent The expression for the energy density of leptons $\varepsilon_{l}$ reads
\begin{equation}
    \varepsilon_{l}= \frac{|q_{l}|B}{4\pi^{2}}\sum_{l}\sum_{\nu}^{\nu_{\mathrm{max}}}\sum_{s}\bigg[k_{F,\nu,s}^{\,l}E_{F}^{\,l}+
\bar{m}_{l}^{2}\ln\bigg|\frac{k_{F,\nu,s}^{\,l}+E_{F}^{\,l}}{\bar{m}_{l}}\bigg|\bigg]\,. \label{ea3}
\end{equation}

The pressures of baryons and leptons are:
\begin{eqnarray}\label{pressurem}
P_{m}&=&\mu_{n}\sum_{b}\rho_{b}^{v}-\varepsilon_{m}, \nonumber \\
P_{l}&=&\sum_{l}\mu_{l}\rho_{l}^{v}-\varepsilon_{l},
\end{eqnarray}

\noindent where the expression of the vector densities $\rho_{b}^{v}$ and $\rho_{l}^{v}$ are given in~(\ref{densities2}) and (\ref{vector_{density_{leptons}}}), respectively. The total energy density and the total pressure of the system can be written by adding the corresponding contributions of the magnetic field:
\begin{equation}\label{et}
    \varepsilon^{\mathrm{H}}=\varepsilon_{m}+\varepsilon_{l}+\frac{\bigg(B\left(\frac{\rho}{\rho_{0}} \right)\bigg)^{2}}{2}\,, \qquad P^{\mathrm{H}}=P_{m}+P_{l}+\frac{\bigg(B\left(\frac{\rho}{\rho_{0}} \right)\bigg)^{2}}{2}
\end{equation}

\subsection{Quark phase under a magnetic field}

For the description of the equation of state (EOS) of quark matter, we consider a (three flavor) quark matter in $\beta$ equilibrium with magnetic fields. We introduce the lagrangian density
\begin{equation}\label{lnjl}
    \mathcal{L}_{Q}=\mathcal{L}_{f}+\mathcal{L}_{l}+\mathcal{L}_{B}, \,
\end{equation}

\noindent where the quark sector is described by the SU(3) version of the Nambu-Jona-Lasinio model (NJL) \cite{hat}, which includes a scalar-pseudoscalar interaction and the t\`{}Hooft six-fermion interaction. The lagrangian density $\mathcal{L}_{l}$ and $\mathcal{L}_{B}$ are given by~(\ref{ll}) and~(\ref{lcm}), respectively. The lagrangian density $\mathcal{L}_{f}$ is defined by
\begin{equation}\label{lf}
    \mathcal{L}_{f}=\bar{\psi}_{f}\left[ \gamma_{\mu}\left( i\partial^{\mu}-q_{f}A^{\mu} \right)-\hat{m}_{c} \right]\psi_{f}+\mathcal{L}_{\mathrm{sym}}+\mathcal{L}_{\mathrm{det}}\,,
\end{equation}

\noindent with
\begin{equation}\label{lsym}
    \mathcal{L}_{\mathrm{sym}}=G\sum_{a=0}^{8}\left[ \left( \bar{\psi}_{f}\lambda_{a}\psi_{f} \right)^{2}
    +\left( \bar{\psi}_{f}i\gamma_{5}\lambda_{a}\psi_{f} \right)^{2} \right]\,,
\end{equation}
\begin{equation}\label{ldet}
    \mathcal{L}_{\mathrm{det}}=-K \left( d_{+}+d_{-}\right)\,,
\end{equation}

\noindent where $G$ and $K$ are coupling constants, $d_{\pm}=\mathrm{det}_{f}\left[ \bar{\psi}_{f}\left( 1\pm\gamma_{5} \right)\psi_{f} \right]$, $\psi_{f}=\left(u,d,s  \right)^{T}$ represents a quark field with three flavors, $\hat{m}_{c}=\mathrm{diag}_{f}\left( m_{u},m_{d},m_{s} \right)$ is the corresponding (current) mass matrix while $q_{f}$ represents the quark electric charge; $\lambda_{0}=\sqrt{2/3}\,I$, where $I$ is the unit matrix in the three flavor space; and $0<\lambda_{a}\leq8$ denote the Gell-Mann matrices. We consider $m_{u}=m_{d}\neq m_{s}$. In the mean-field approximation the lagrangian density~(\ref{lf}) can be written as \cite{mene}
\begin{eqnarray}\label{lf_mfa}
  \mathcal{L}_{f}^{\mathrm{MFA}} &=& \bar{\psi}_{f}\left[ \gamma_{\mu}\left( i\partial^{\mu}-q_{f}A^{\mu} \right)-\hat{M}\right]\psi_{f}\nonumber \\
    & & -2G\left( \phi_{u}^{2}+\phi_{d}^{2}+\phi_{s}^{2} \right)+4K\phi_{u}\phi_{d}\phi_{s}\,,
\end{eqnarray}

\noindent where $\hat{M}$ is a diagonal matrix with elements defined by the
effective quark masses
\begin{equation}\label{eqm}
    M_{i}=m_{i}-4G\phi_{i}+2K\phi_{j}\phi_{k}
\end{equation}

\noindent with $\left( i,j,k \right)$ being some permutation of $\left( u,d,s \right)$.

Now, we need to evaluate the grand-canonical thermodynamical potential for the three-flavor quark sector, which can be written as
$\Omega_{f}=-P_{f}=\varepsilon_{f}-\sum_{f}\mu_{f}\rho_{f}-\Omega_{0}$, where $P_{f}$ represents the pressure, $\varepsilon_{f}$ the energy density, $\mu_{f}$ the chemical potential and $\Omega_{0}$ ensures that $\Omega_{f}=0$ in the vacuum. In the mean-field approximation the pressure can be written as
\begin{equation}\label{p_mfa}
    P_{f}=\theta_{u}+\theta_{d}+\theta_{s}-2G\left( \phi_{u}^{2}+\phi_{d}^{2}+\phi_{s}^{2} \right)+4K\phi_{u}\phi_{d}\phi_{s}\,.
\end{equation}

So, to determine the EOS for the SU(3) NJL model at finite density and in the presence of a magnetic field we need to know the condensates $\phi_{f}$ and the contribution from the gas of quasiparticles $\theta_{f}$. Both quantities have been evaluated with great detail in Refs.~\cite{mene2,mene}. For this model we split the degeneracy of each quark into the spin degeneracy and color degeneracy $N_c$. The difference now is that both spin projections contribute for Landau levels $\nu>0$, but only one of them contributes for $\nu=0$. The contribution from the gas of quasiparticles for each flavor $\theta_{f}=\left(\theta^{\mathrm{vac}}_{f}+\theta^{\mathrm{mag}}_{f}+\theta^{\mathrm{med}}_{f}\right)_{M_{f}}$ contains $3$ different contributions: the vacuum, the magnetic and the medium one given by
\begin{equation}
\theta^{\mathrm{vac}}_{f}=- \frac{N_c}{8\pi^2} \left \{ M_f^4 \ln \left[\frac{(\Lambda+ \epsilon_\Lambda)}{M_f} \right]
-\epsilon_\Lambda \, \Lambda\left(\Lambda^2 +  \epsilon_\Lambda^2 \right ) \right \} \mbox{,}
\end{equation}
\begin{equation}
\theta^{\mathrm{mag}}_{f}= \frac {N_{c} (|q_{f}|B)^2}{2 \pi^2} \left [ \zeta^{(1,0)}(-1,x_{f}) -  \frac {1}{2}( x_{f}^{2} - x_{f}) \ln x_{f} +\frac {x_{f}^{2}}{4} \right ] \mbox{,}
\end{equation}
\begin{eqnarray}
\theta^{\mathrm{med}}_{f}&=&\sum_\nu\frac{\alpha_\nu N_c|q_{f}|B }{4\pi^2} \left[ \mu_{f}\sqrt{\mu_{f}^{2}-s_{f}(\nu,B)^{2}} \right.\nonumber\\
& &\left . -s_{f}(\nu,B)^{2}\ln\left( \frac{\mu_{f}+\sqrt{\mu_{f}^{2}-s_{f}(\nu,B)^{2}}}{s_{f}(\nu,B)} \right) \right]  \mbox{,}
\label{PmuB}
\end{eqnarray}

\noindent where $s_{f}(\nu,B)=\sqrt{M_{f}^2+2|q_{f}|B\nu}$ is the constituent mass of each quark modified by the magnetic field, $\epsilon_\Lambda=\sqrt{\Lambda^2 + M_{f}^{2}}$ with $\Lambda$ representing a non covariant ultra violet cut off \cite{ebert},  $x_f = M_{f}^{2}/(2 |q_{f}| B)$ and $\zeta^{(1,0)}(-1,x_f)= d\zeta(z,x_f)/dz|_{z=-1}$ with $\zeta(z,x_f)$ being the Riemann-Hurwitz zeta function.

Each of the quark condensates, $\phi_{f}=\langle{\bar \psi}_{f} \psi_{f}\rangle=(\phi_{f}^{\mathrm{vac}}+\phi_{f}^{\mathrm{mag}}+\phi_{f}^{\mathrm{med}})_{M_{f}}$ also contains $3$ different contributions: the vacuum, the magnetic and the medium one given by ~\cite{mene2}
\begin{eqnarray}
\phi_{f}^{\mathrm{vac}} &=& -\frac{N_c  M_{f}}{2\pi^{2}} \left[\Lambda \epsilon_\Lambda-{M_{f}^{2}}\ln \left ( \frac{\Lambda+ \epsilon_\Lambda}{{M_{f} }} \right ) \right ] \mbox{,}
\end{eqnarray}
\begin{eqnarray}
\phi_{f}^{\mathrm{mag}}&=& -\frac{N_{c} M_{f}|q_{f}| B}{2\pi^{2}}\left [ \ln \Gamma(x_{f}) -\frac{1}{2} \ln (2\pi) \right . \nonumber \\
& &+ \left . x_{f} -\frac{1}{2} \left ( 2 x_{f}-1 \right )\ln (x_{f}) \right ] \mbox{,}
\end{eqnarray}
\begin{eqnarray}\label{MmuB}
\phi_{f}^{\mathrm{med}}&=&\sum_{\nu}\frac{\alpha_\nu N_{c} M_{f} |q_{f}| B}{2\pi^2}\nonumber\\
& & \times \left[ \ln\left( \frac{\mu_{f}+\sqrt{\mu_{f}^{2}-s_{f}(\nu,B)^{2}}}{s_{f}(\nu,B)} \right) \right]\mbox{.}
\end{eqnarray}

The quark contribution to the energy density is
\begin{equation}\label{energy_f}
    \varepsilon_{f}=-P_{f}+\sum_{f}\mu_{f}\rho_{f}+\Omega_{0}\,,
\end{equation}

\noindent where the density $\rho_{f}$ corresponds to each different flavor and is given by
\begin{equation}\label{rho_f}
    \rho_{f}=\sum_{\nu}\frac{\alpha_{\nu}N_{c}|q_{f}|B}{2\pi^{2}}\,k_{F,f} \mbox{,}
\end{equation}

\noindent with $k_{F,f}=\sqrt{\mu_{f}^{2}-s_{f}(\nu,B)^{2}}$.

The leptonic contribution for the pressure reads
\begin{eqnarray}\label{pl}
P_{l}&=&\sum_{l}\sum_\nu\frac{\alpha_\nu |q_{l}|B }{4\pi^2} \left[ \mu_{l}\sqrt{\mu_{l}^{2}-s_{l}(\nu,B)^{2}} \right.\nonumber\\
& &\left . -s_{l}(\nu,B)^{2}\ln\left( \frac{\mu_{l}+\sqrt{\mu_{l}^{2}-s_{l}(\nu,B)^{2}}}{s_{l}(\nu,B)} \right) \right]  \mbox{,}
\end{eqnarray}

\noindent and finally the vector density and energy density for leptons are given by the equations (\ref{vector_{density_{leptons}}}) and (\ref{ea3}), respectively. Therefore, the total energy density and the total pressure of the system are given by adding the corresponding contribution of the magnetic field
\begin{equation}\label{etq}
    \varepsilon^{\mathrm{Q}}=\varepsilon_{f}+\varepsilon_{l}+\frac{\left(B\left( \frac{\rho}{\rho_{0}} \right)\right)^{2}}{2}\,, \qquad P^{\mathrm{Q}}=P_{f}+P_{l}+\frac{\left(B\left( \frac{\rho}{\rho_{0}} \right)\right)^{2}}{2}\,.
\end{equation}

The parameter sets of the NJL model used in the present work are given in
TABLE \ref{pnjl}.
\begin{table}[h]
\centering
\begin{tabular}{|c|c|c|c|c|c|}
  \hline
   & $\Lambda$ & $G\Lambda^2$ & $K\Lambda^5$ & $m_{u,d}$ & $m_{s}$ \\
      Parameter set     &    (MeV)     &          &   &  (MeV)   &  (MeV)  \\ \hline
  SU(3) HK \cite{Buballa,hat2}& 631.4 & 1.835 & 9.29 & 5.5 & 135.7 \\ \hline
  SU(3) RKH\cite{Buballa,reh} & 602.3 & 1.835 & 12.36 & 5.5 & 140.7 \\
  \hline
\end{tabular}
\caption{\label{pnjl} Parameter sets for the NJL SU(3) model.}
\end{table}

\subsection{Hybrid star}
\label{sec:C}
There are two ways of constructing a hybrid star, one with a mixed phase and another without a mixed phase (hadron and quark phases are in direct contact). In the first case, neutron and electron chemical potentials are continuous throughout the stellar matter, based on the standard thermodynamical rules for phase coexistence known as Gibbs condition \cite{glen,vosk,maru,marce}. In the second case, the electron chemical potential suffers a discontinuity because only the neutron chemical potential is imposed to be continuous. The condition underlying the fact that only a single chemical potential is common to both phases is known as Maxwell condition. Recently, some authors calculated macroscopic quantities as radii and masses for hybrid stars with and without mixed phase and they concluded that the differences were not relevant~\cite{vosk,maru,marce}. Inspired by these results, in the present work we choose the simpler construction for a hybrid star which is based on the Maxwell condition.

For the construction of a hybrid star with the Maxwell condition, we just need to find the point where
\begin{equation}\label{MC}
    \mu_{n}^{\mathrm{H}}=\mu_{n}^{\mathrm{Q}} \qquad \mathrm{and} \qquad P^{\mathrm{H}}=P^{\mathrm{Q}}\,.
\end{equation}

%--------------------------------

\noindent To construct a hybrid star we consider a system constituted by $8$ baryons in the hadron phase and $3$ quarks in the quark phase. For the EOS of the hadronic phase we use equation~(\ref{et}) with $\kappa_{b}=0$ in the equations (\ref{ll2}), (\ref{ea1}) and (\ref{ea2}) (i.e. without anomalous magnetic moment) and for the EOS of the quark phase we use equation~(\ref{etq}).

%-------------------------------

\section{Results}

In the sequel we consider two different systems under a strong magnetic field: (A) baryonic, and (B) hybrid matters. In both cases the effects of strong magnetic fields on the macroscopic properties of compact stars were obtained from the integration of the Tolman-Oppenheimer-Volkoff (TOV) equations~\cite{tov}, using as input the EOS obtained from subsections \ref{sec:A} and \ref{sec:C}
for baryonic and hybrid matters, respectively.

\begin{figure}[ht]
\begin{center}
\includegraphics[angle=270,width=0.45\linewidth]{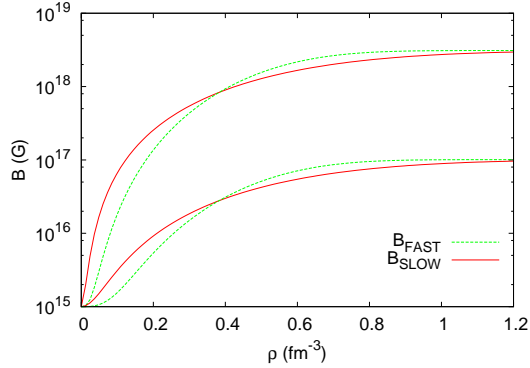}	
\end{center}
\caption{ Variable density dependent magnetic fields for $B_{0}=10^{17}$ G (lower curves) and $B=3.1\times 10^{18}$ G (upper curves), for FAST (green line) and SLOW (red line) decays.}
 \label{FIG-A}
\end{figure}

We assume that the density-dependent magnetic field $B$ in the EOS is given
by~\cite{bandy,mene,lopes,panda,mao}:
\begin{equation}\label{bdd}
    B\left( \frac{\rho}{\rho_{0}} \right)=B_{\mathrm{surf}}+B_{0}\left\{ 1-\mathrm{exp}\left[
    -\beta\left( \frac{\rho}{\rho_{0}} \right)^{\gamma} \right] \right\}\,,
\end{equation}

\noindent where $\rho=\sum_{b}\rho_{b}^{v}$ is the baryon density, $\rho_{0}$ is the saturation density, $B_{\mathrm{surf}}$ is the magnetic field on the surface taken equal to $10^{15}~G$ in agreement with observational values and $B_{0}$ is the magnetic field for larger densities. The remaining parameters $\beta$ and $\gamma$ are chosen to reproduce two behaviors of the magnetic field: a fast decay with  $\gamma=3.00$ and $\beta=0.02$, and a slow decay with $\gamma=2.00$ and $\beta=0.05$ whose curves can be seen in Fig.~\ref{FIG-A}.

In \cite{sinha}, different profiles for the density dependence of the magnetic
field were studied in the context of hadronic stars and the authors concluded
that the equation of state is insensitive to magnetic fields lower than
$10^{18}$ G, a behaviour already observed in \cite{mene,mene2,lopes}
for different models. Moreover, they found that for magnetic fields higher than
$10^{18}$ G, matter becomes unstable due to the increase of anisotropic effects
on the pressure. Taking into account that $10^{18}$ G seems to establish
two different boundaries and the anisotropic effects around
$3.1 \times 10^{18}$ G is small \cite{nosso_2013}, resulting in a small error
in the stellar masses if the TOV equations are used \cite{mallick}, we next
use two values for
the magnetic field, namely $10^{17}$ G and  $3.1 \times 10^{18}$ G and include
the effects of the AMM.

\subsection{Baryonic matter}

\begin{figure}[ht]
\begin{center}
\begin{tabular}{ll}
\includegraphics[angle=270,width=0.50\linewidth]{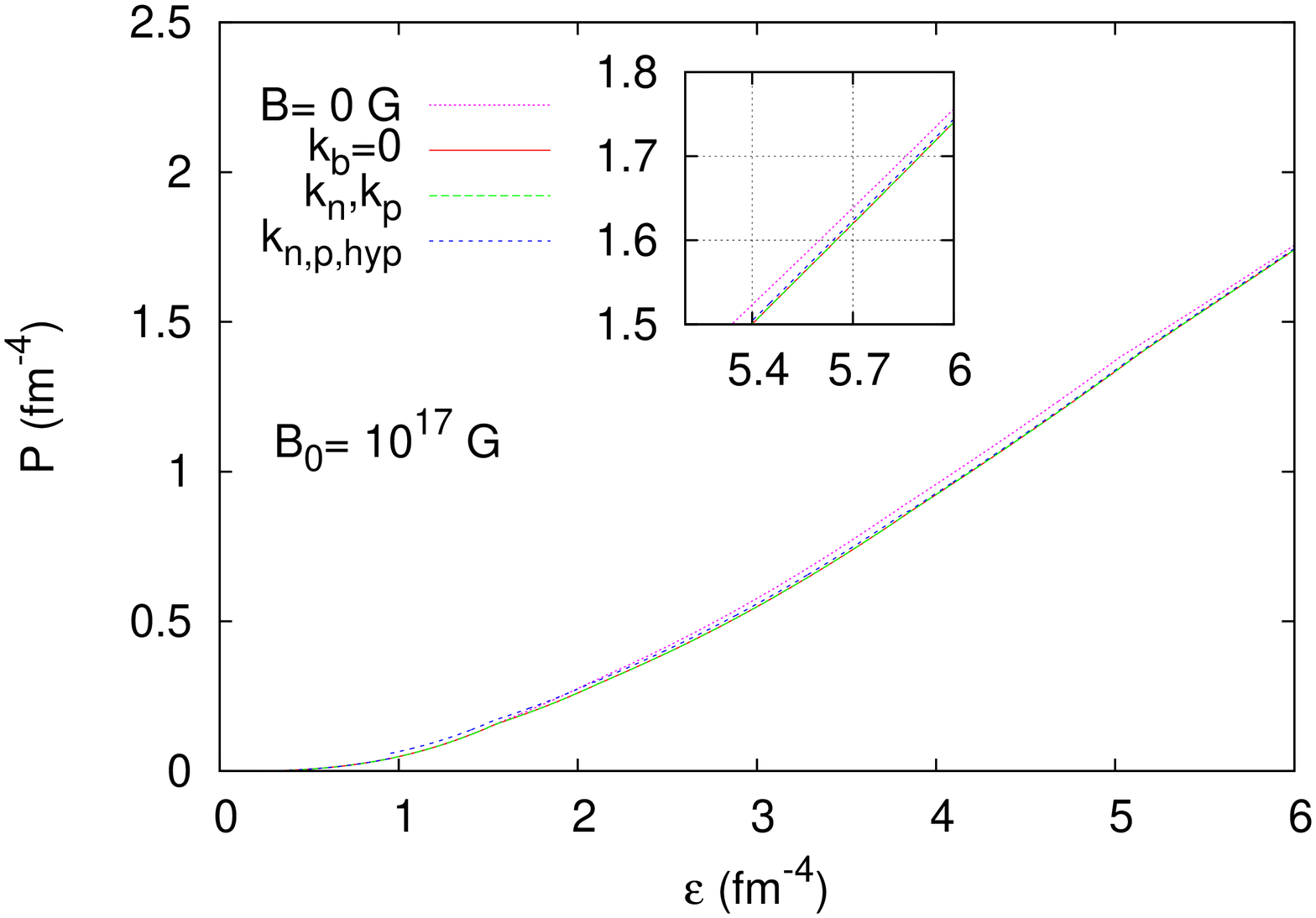}&
\includegraphics[angle=270,width=0.50\linewidth]{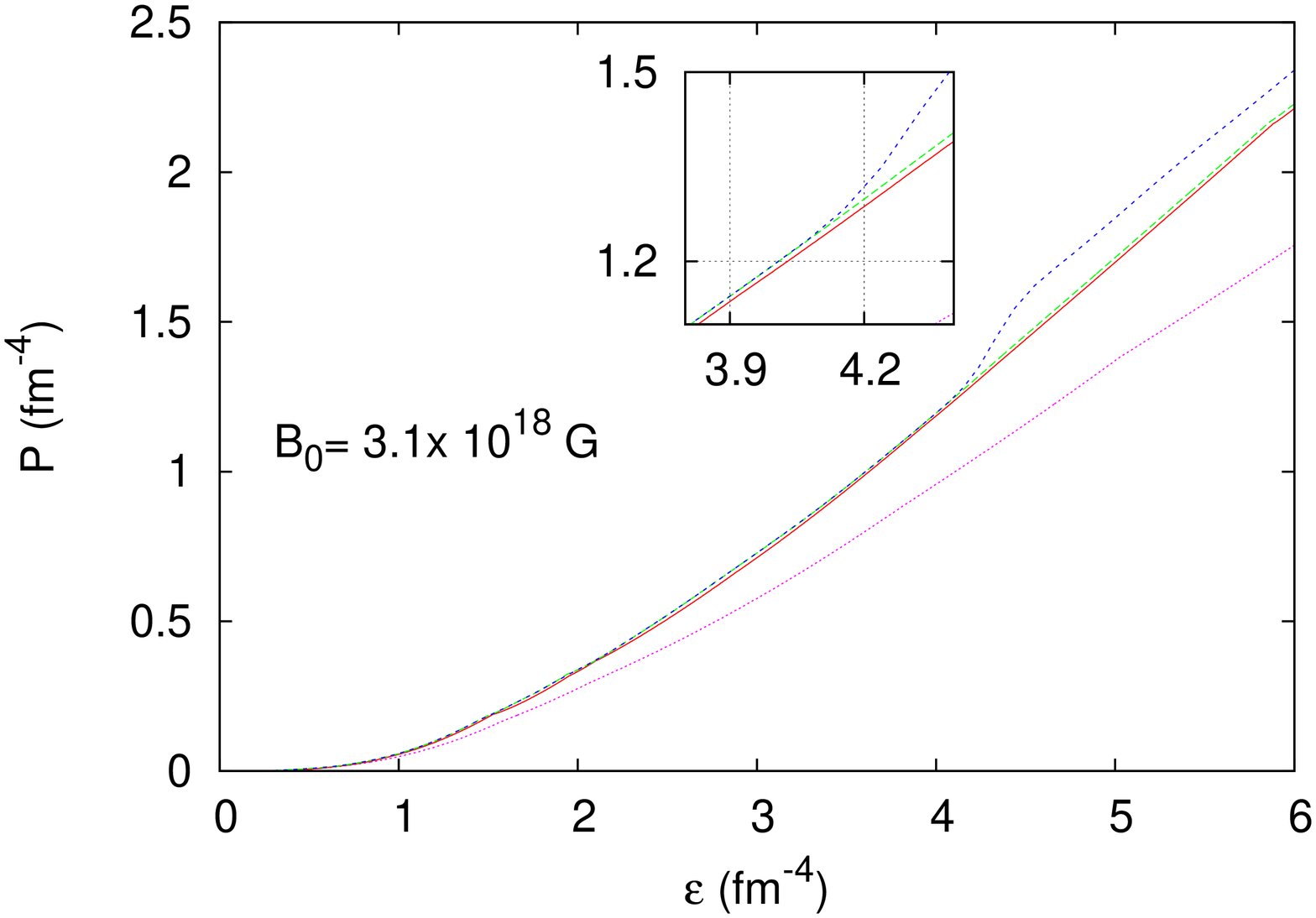} \\
\includegraphics[angle=270,width=0.50\linewidth]{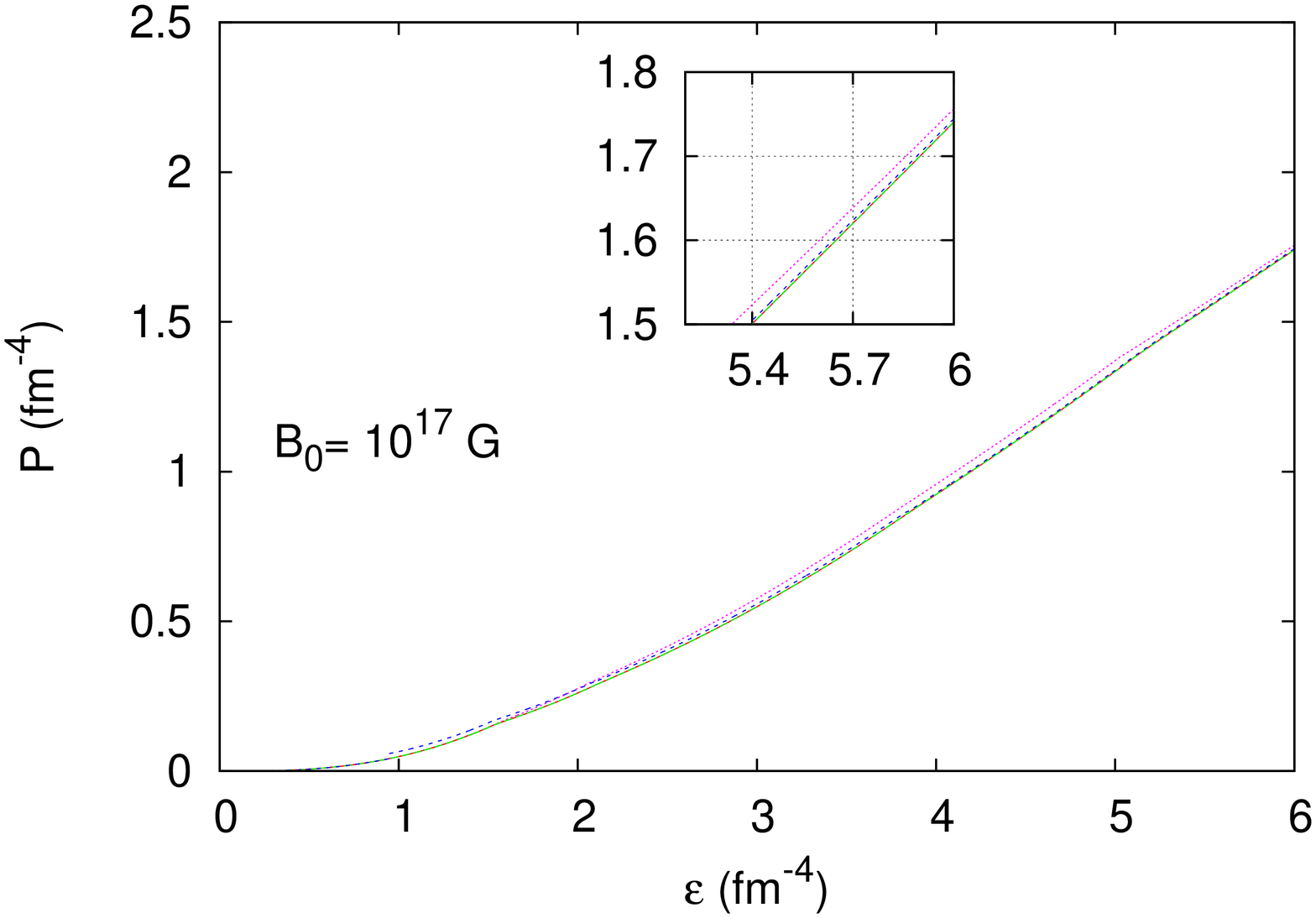}&
\includegraphics[angle=270,width=0.50\linewidth]{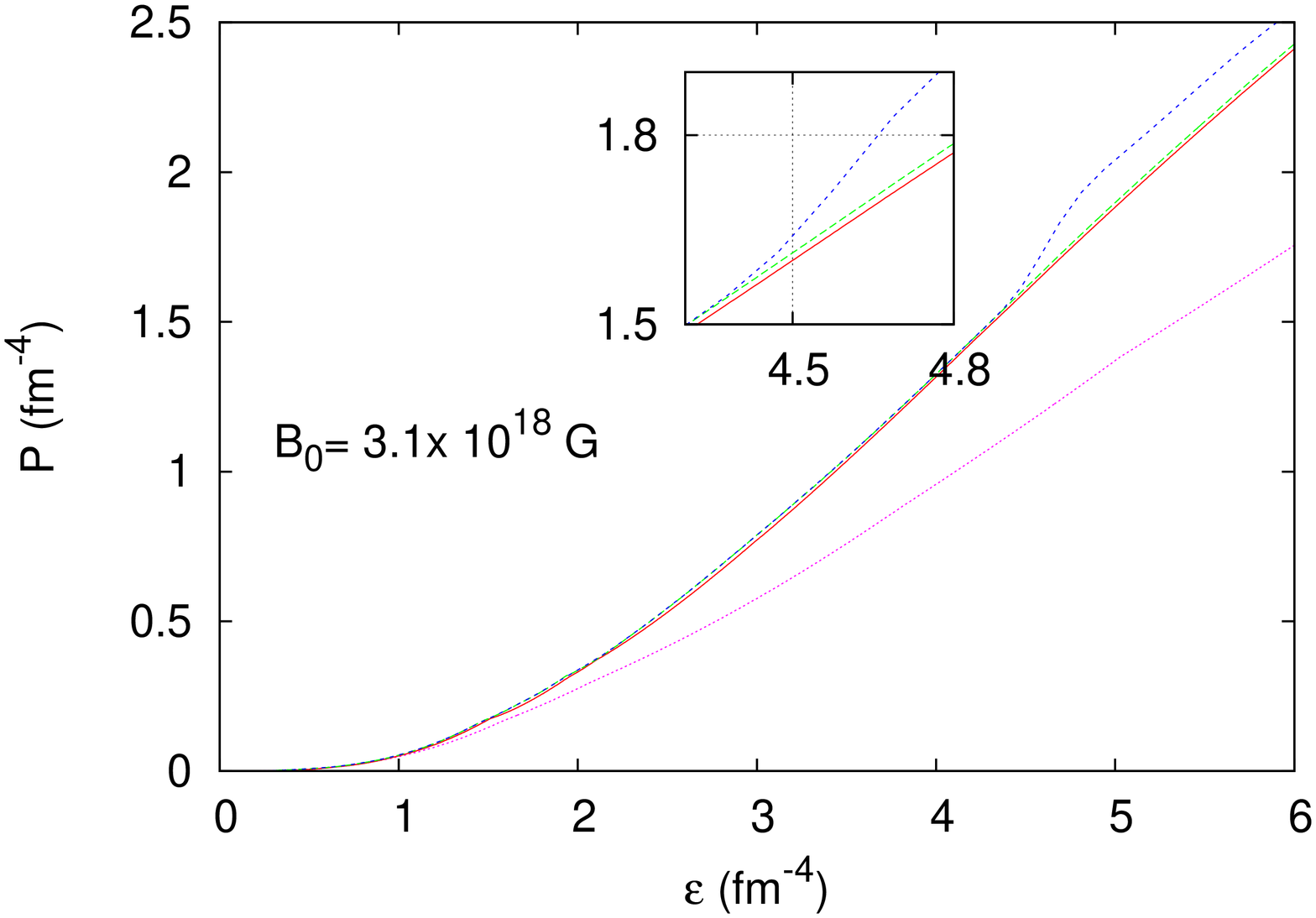} \\
\end{tabular}
\end{center}
\caption{Equations of State for hadronic matter, with the inclusion of the baryonic octet. Three cases for the inclusion of the anomalous magnetic moments are considered, for slow (upper panels) and fast (lower panels) decays, and for $B_{0}=10^{17}~G$ (left panels) and $B_{0}=3.1 \times 10^{18}~G$ (right panels).}
\label{FIG-B}
\end{figure}

In Fig.~\ref{FIG-B} we show the equation of state for hadronic matter under the influence of $B_{0}=10^{17}~G$ (left panels) and $B_{0}=3.1 \times 10^{18}~G$ (right panels) magnetic fields, and with three possible conditions for the inclusion of the anomalous magnetic moment, ($k_{b}=0$) for no corrections, ($k_{n}$) and ($k_{p}$) for the inclusion of the neutron and proton anomalous magnetic moments  and ($k_{n,p,hyp}$) for the inclusion of the corrections for all the baryons, both for slow (upper panels)
and fast (lower panels) decays. We see no great difference in any of the cases studied for $B_{0}=10^{17}~G$. As expected, they practically coincide with the non-magnetized curve  (in red), as can be seen in the zoomed boxes.

 At $B_{0}=3.1 \times 10^{18}~G$ we notice the stiffening caused by the larger magnetic field applied, on both fast and slow cases. On the zoomed boxes it is possible to notice the stiffening effects of the inclusion of the corrections due to the magnetic moments, even when only $\kappa_{n}$ and $\kappa_{p}$ are included. This effect is stronger for higher energy densities, which coincides qualitatively with the the effects caused on nucleonic matter found in \cite{Stellar-matter-with}. The effect of the inclusion of the anomalous magnetic moment of all the hyperons only becomes evident at higher values of the energy density.

\begin{figure}[ht]
\begin{center}
\begin{tabular}{ll}
\includegraphics[angle=270,width=0.50\linewidth]{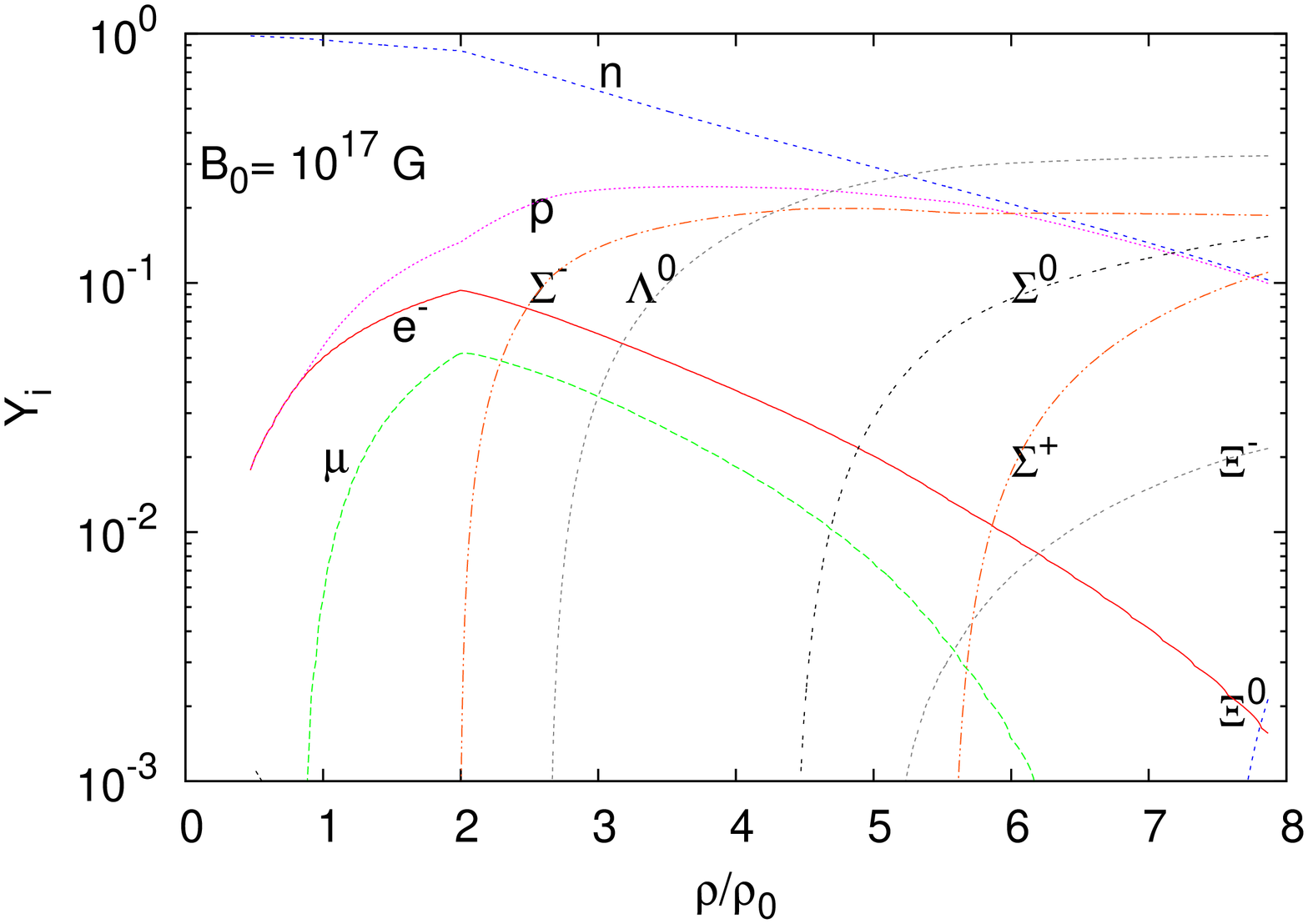} &
\includegraphics[angle=270,width=0.50\linewidth]{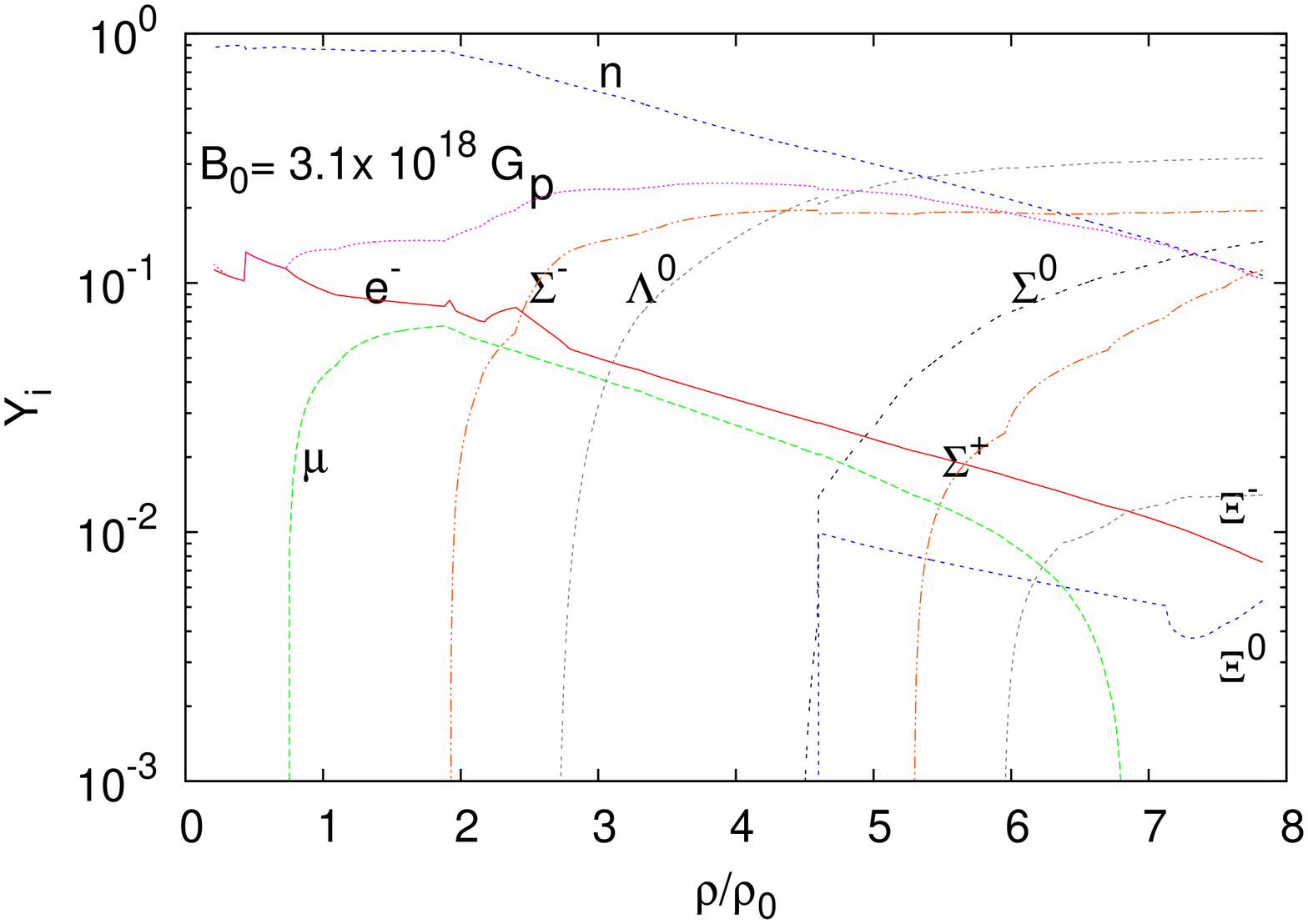} \\
\end{tabular}
\end{center}
\caption{Particle fractions for hadronic matter, with the inclusion of the anomalous magnetic moment for all the baryonic octet, for $B_{0}=10^{17}~G$ (left panel) and $B_{0}=3.1 \times 10^{18}~G$ (right panel).}
\label{FIG-C}
\end{figure}

In Fig.~\ref{FIG-C} we present the particle fractions for hadronic matter with the inclusion of the anomalous magnetic moment of all particles for $B_{0}=10^{17}~G$ (left panel) and $B_{0}=3.1 \times 10^{18}~G$ (right panel).
Comparing the two graphs, we see different behaviors of some abundances caused by the increase in the intensity of the magnetic field, like the kinks produced on the populations of charged particles, due to the filling of Landau levels.

\begin{figure}[ht]
\begin{center}
\begin{tabular}{ll}
\includegraphics[angle=270,width=0.50\linewidth]{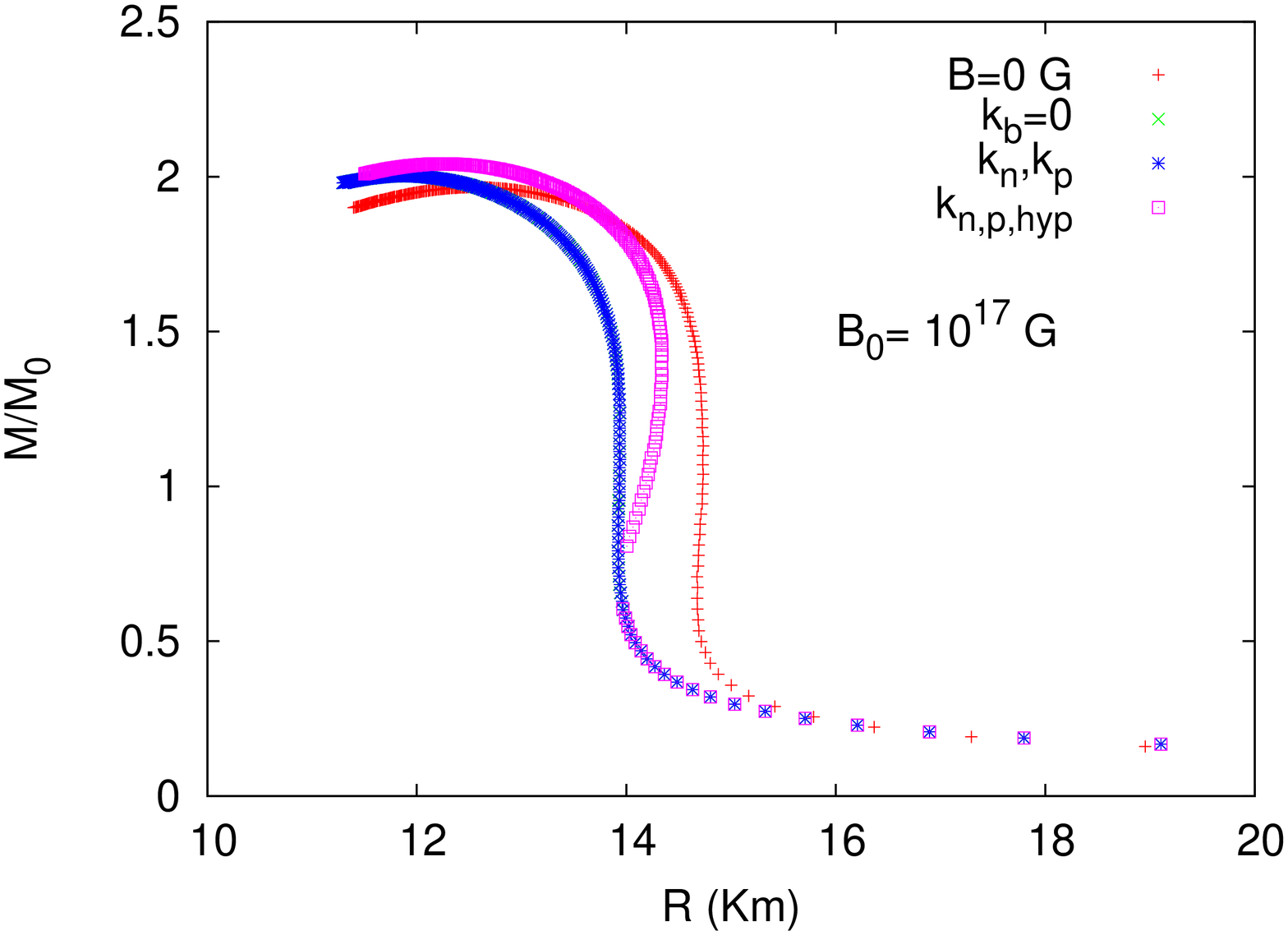} &
\includegraphics[angle=270,width=0.50\linewidth]{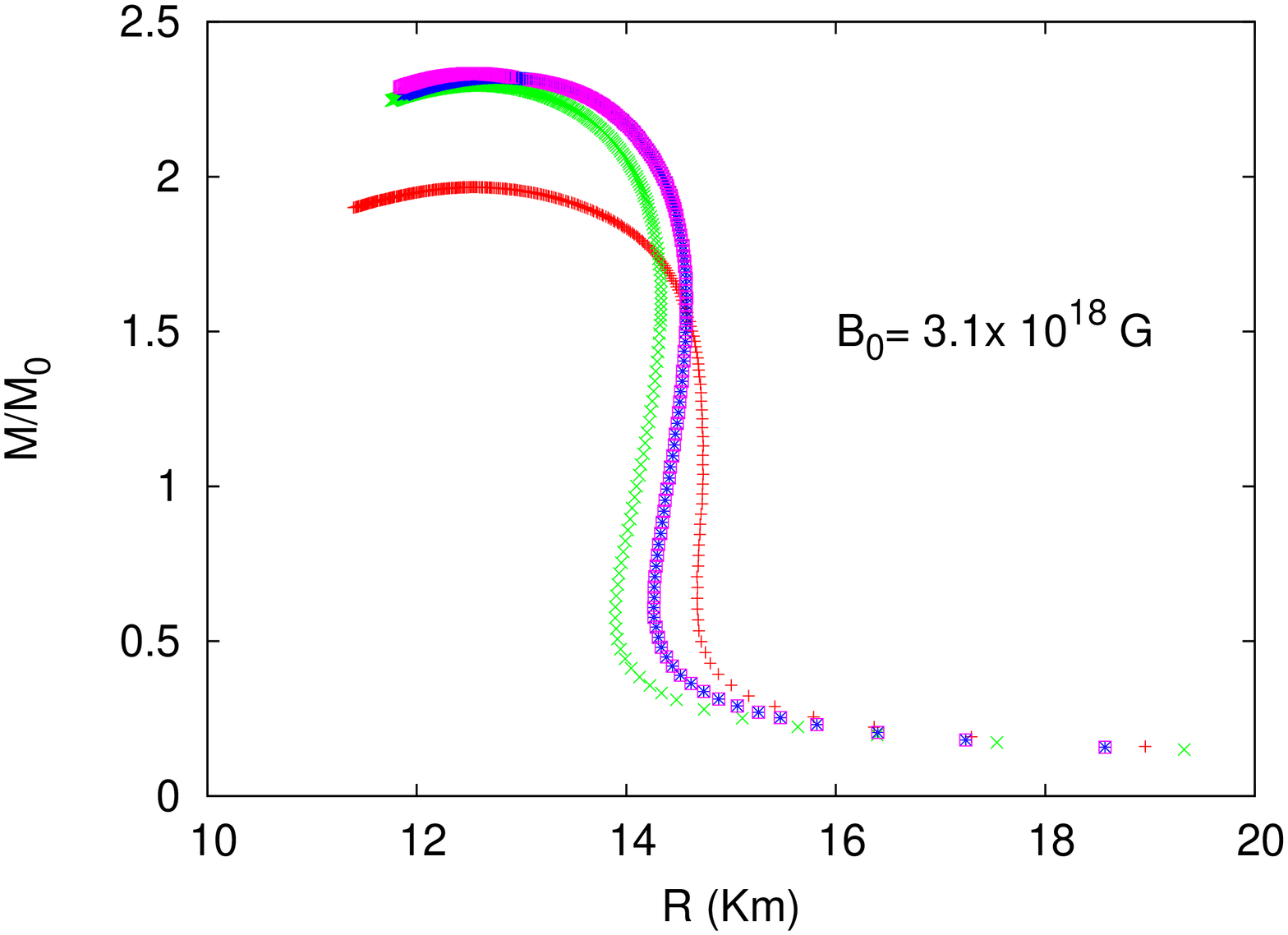}\\
\includegraphics[angle=270,width=0.50\linewidth]{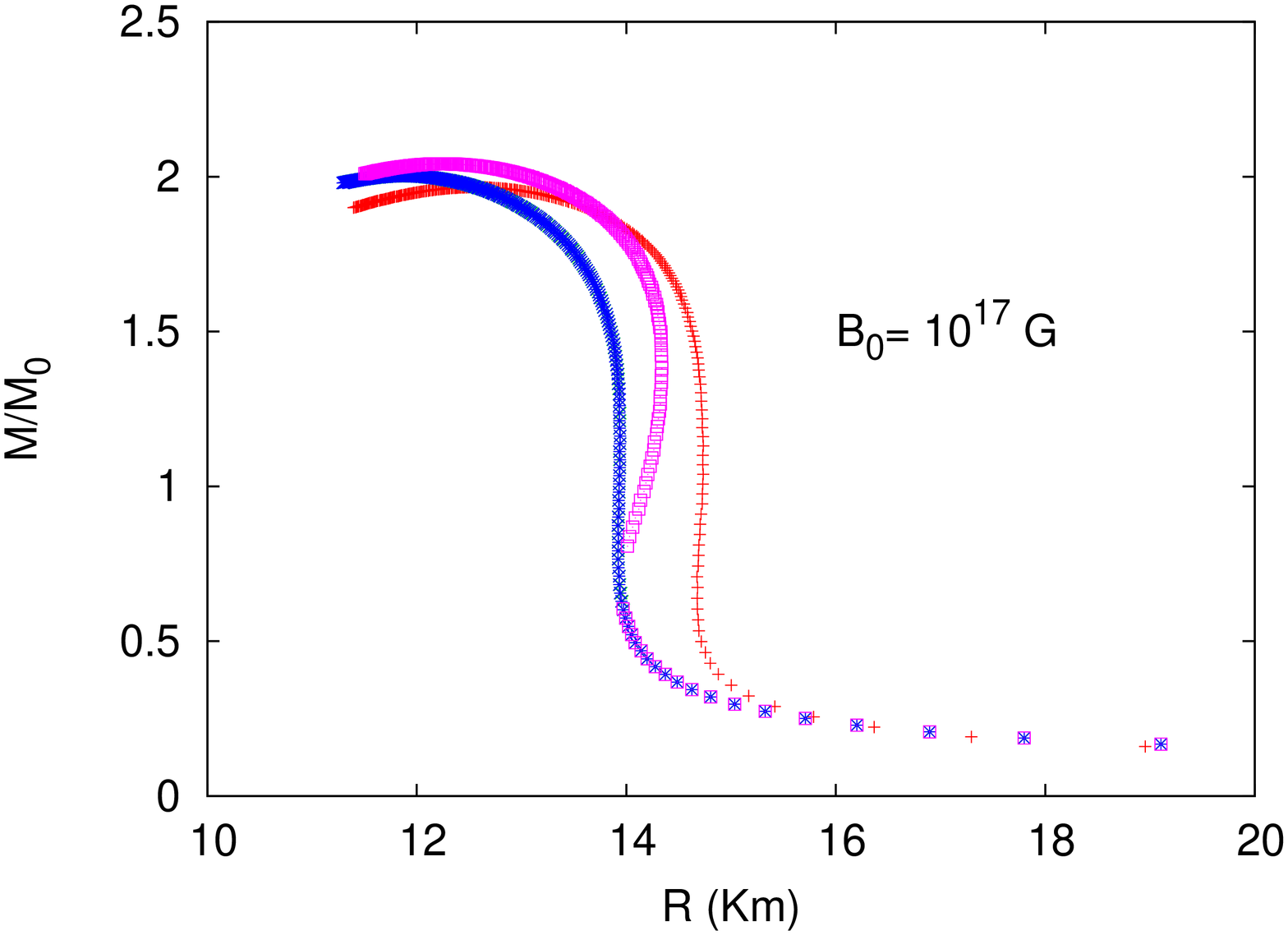} &
\includegraphics[angle=270,width=0.50\linewidth]{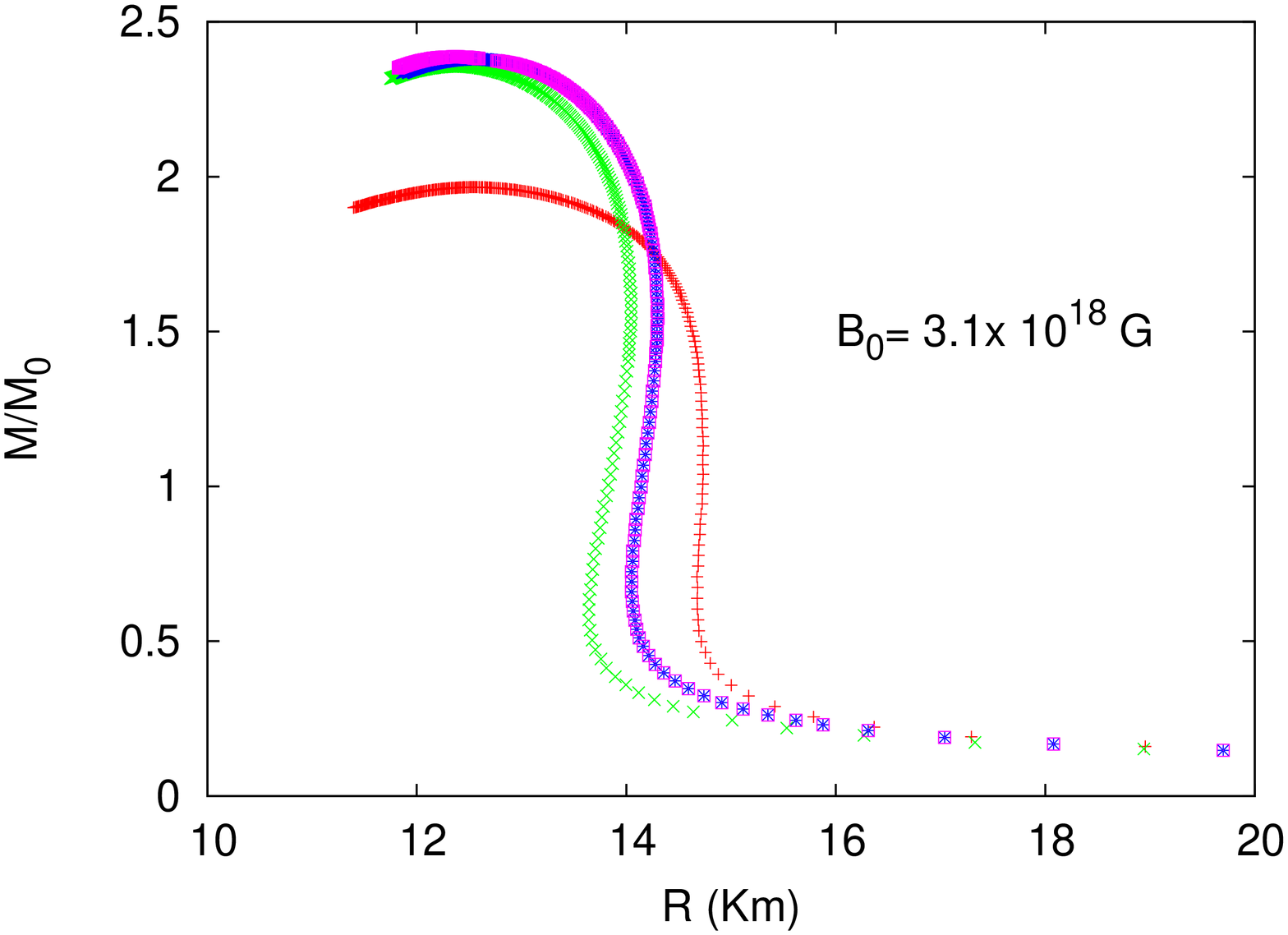}\\
\end{tabular}
\end{center}
\caption{Mass-radius curves for hadronic matter, with the inclusion of the baryonic octet. Three cases of anomalous magnetic moments are considered, $k_{b}=0$ without magnetic moment corrections, $k_{n}$ and $k_{p}$ with corrections for neutrons and protons and $k_{n,p,hyp}$ with the anomalous magnetic moment of all baryons. For $B_{0}=10^{17}~G$ (left panel) and $B_{0}=3.1 \times 10^{18}~G$ (right panel), for slow (upper panels) and fast (lower panels) decays. }
\label{FIG-D}
\end{figure}

In Fig.~\ref{FIG-D} we plot the mass-radius relation of hadronic matter under the influence of $B_{0}=10^{17}~G$ (left panel) and $B_{0}=3.1 \times 10^{18}~G$ (right panel) magnetic fields, and with the three possible conditions for the inclusion of the magnetic moment corrections presented before, both for slow (upper panels) and fast (lower panels) decays. The tails of the hadronic matter were obtained with the insertion of the BPS EOS \cite{bps}. On the left as expected, the curves for $B_{0}=10^{17}~G$ present maximum masses and radii that do not differ greatly from those found for the non-magnetized EOS (in red), used for comparison,
as already expected from previous results in the literature \cite{bro2}.
On the right, for the upper and lower panels, due to the stiffening of the curves for $B_{0}=3.1\times 10^{18}~G$, caused by the stronger magnetic field, the effects caused by the different behaviors on the decay of the equation (\ref{bdd}), become more evident, and the extensions of these effects on equation (\ref{et}), from slow to fast decay, generate higher maximum masses and lower radii, for all of the anomalous magnetic moment conditions considered. This can be seen in TABLE \ref{table4}.

\newpage

\begin{sidewaystable}
\begin{tabular}{|c|c|c|c|c|c|c|c|c|c|c|c|c|c|}
  \hline
Magnetic Field &AMM&\multicolumn{6}{ |c| }{FAST}&\multicolumn{6}{ |c| }{SLOW} \\
  \hline
&&$M_{max}$& R & $\varepsilon_{c}$ & $\mu_{n}(\varepsilon_{c})$ & $\mu_{e}(\varepsilon_{c})$ & R($M=1.4~M_{0}$) &
$M_{max}$&R  & $\varepsilon_{c}$& $\mu_{n}(\varepsilon_{c})$ & $\mu_{e}(\varepsilon_{c})$& R($M=1.4~M_{0}$)    \\ \cline{3-14}
&& $(M_{0})$& (Km)& (fm$^{-4}$) & (MeV) & (MeV) & (Km) & $(M_{0})$& (Km) & (fm$^{-4}$) & (MeV) & (MeV) & (Km) \\ \cline{3-14}
  \hline
$B=0~G$ & & 1.97 & 12.55& 5.29 & 1417.9 & 93.9 & 14.69 & 1.97 & 12.55 & 5.29&
1417.9 & 93.9 &  14.69\\
\hline
& $\kappa_{b}=0$  & 2.00& 11.87 & 5.93& 1577.5 & 122.1 & 13.90&2.00& 11.87 & 5.93& 1577.5 & 122.1 & 13.90\\ \cline{2-14}
$B_{0}=10^{17}~G$   & $\kappa_{n,p}$ & 2.00 & 11.87 & 5.93& 1577.5 & 122.1 &13.91& 2.00& 11.87 & 5.93& 1577.5 & 122.1 &13.91\\ \cline{2-14}
& $\kappa_{n,p,hyp}$& 2.04& 12.24& 5.56& 1549.3 & 131.4 & 14.34& 2.04& 12.24 & 5.56& 1549.3 & 131.4 & 14.33\\ \cline{2-14}
  \hline
& $\kappa_{b}=0$ & 2.36& 12.37& 5.27& 1427.3 & 150.2 &14.02& 2.29& 12.58 & 5.11&
1446.1 & 150.2 &14.29\\ \cline{2-14}
$B_{0}=3.1\times 10^{18}~G$ & $\kappa_{n,p}$ & 2.38& 12.53& 5.16 & 1417.9 & 159.6 &14.23& 2.32& 12.77 & 4.97 & 1436.7 & 150.2 &14.55\\ \cline{2-14}
& $\kappa_{n,p,hyp}$& 2.39& 12.38 & 5.34& 1436.7 & 150.2 &14.23&  2.33& 12.54 & 5.25& 1464.8 & 150.2 &14.55\\ \cline{2-14}
  \hline
 \end{tabular}
 \caption{Maximum masses and related radii and central energy densities for hadronic matter. $\mu_{n}(\varepsilon_{c})$  and $\mu_{e}(\varepsilon_{c})$ are the chemical potentials for neutron and electron at the central energy density $\varepsilon_{c}$,  R($M=1.4~M_{0}$) is the radius for a $M=1.4~M_{0}$ for this configurations. }
\label{table4}
\end{sidewaystable}
\newpage

It is well known that the inclusion of the hyperons softens the EOS, reducing the maximum stellar mass. However,
for high values of the magnetic fields, we see from \ref{table4}
that the progressive inclusion of the anomalous magnetic moment stiffens the EOS, first with the curves with only the neutron ($k_{n}$) and proton ($k_{p}$) corrections and then with the AMM of all baryons ($k_{n,p,hyp}$), causing the increase of the maximum mass. This happens both with the FAST and SLOW cases.
For the lower value of the magnetic field, the macroscopic results depend
very little on  the inclusion of the AMM, as they also depend only slightly
on $B$.
For the higher value of the magnetic field, when we compare FAST and SLOW cases, we notice that independently of the AMM condition, the maximum masses of the FAST cases are always larger than those of the respective SLOW case.
We attribute this to the greater stiffness of the EOS caused by the faster decay in equation (\ref{bdd}). Notice, however, that this behaviour
depends on the strenght of the magnetic field in the core, as discussed
in \cite{sinha,panda}.

Comparing our results with \cite{lopes} we confirm that the inclusion of low magnetic fields, of the order of $B_{0}=10^{17}~G$ do not produce any significant effect neither on the EOS nor on the particle fractions. No nozzles are noticed, because due to the low magnetic field there are several Landau levels to be filled, even at low densities. Still comparing our results for $B_{0}=10^{17}~G$, we found for our parametrization without the anomalous magnetic moment corrections ($k_{b}=0$) a maximum mass $M_{max}=2.00~M_{0}$ and a radius $R=11.87$~Km compatible with the $M_{max}=2.01~M_{0}$ and $R=11.86$~Km found in ref. \cite{lopes}.

When we compare our results for $B_{0}=3.1\times 10^{18}~G$ with those in \cite{lopes} we confirm the nozzles on the particle fractions at lower densities,
related to the van Alphen oscillations related
to the creation of a new Landau levels, and a behavior close to the continuous at higher densities, due to the higher number of filled Landau levels. On the  other hand, we also find some different results, mainly because of the choice of the decay parameters of equation (\ref{bdd}). For instance, our results for the maximum mass of the EOS with no magnetic moment corrections ($k_{b}=0$) are $M_{max}=2.36~M_{0}$ and $R=12.37$~Km. In ref. \cite{lopes}, for $\beta=6.5\times 10^{-3}$ and $\gamma=3.5$, the authors obtained
$M_{max}=2.22~M_{0}$ and $R=11.80$~Km, which corroborates the conclusions that
the choice of parameters in the density dependent magnetic field influences the macroscopic properties of the stars.
We can observe that the central energy densities attained do not show a well
established pattern: for low magnetic fields, they decrease when all AMM are
included and for high magnetic fields they oscillate when the AMM are added.
As already mentioned in the Introduction, we can see that baryonic chemical
potentials higher than 1500 MeV are not reached for strong magnetic fields,
independently of the chosen decay rate.
For the sake of completeness, we have also added the radii
results for canonical
1.4 M$_\odot$ stars, where we can see that they  are quite large for our
choice of magnetic field decay rates.

\subsection{Hybrid stars}

Now we turn our attention to hybrid stars.
To study the effects of strong magnetic fields on the macroscopic properties of hybrid stars without mixed phase (Maxwell condition) we also assume the density-dependent magnetic field given in Eq.(\ref{bdd}). In this case, we choose the same two sets of values for the parameters $\beta$ and $\gamma$, a fast varying field defined by $\beta=0.02$ and $\gamma=3.00$ and a slowly varying field with $\beta=0.05$ and $\gamma=2.00$ as in the last section.
\begin{figure}
\begin{center}
\begin{tabular}{ll}
\includegraphics[angle=270,width=0.50\linewidth]{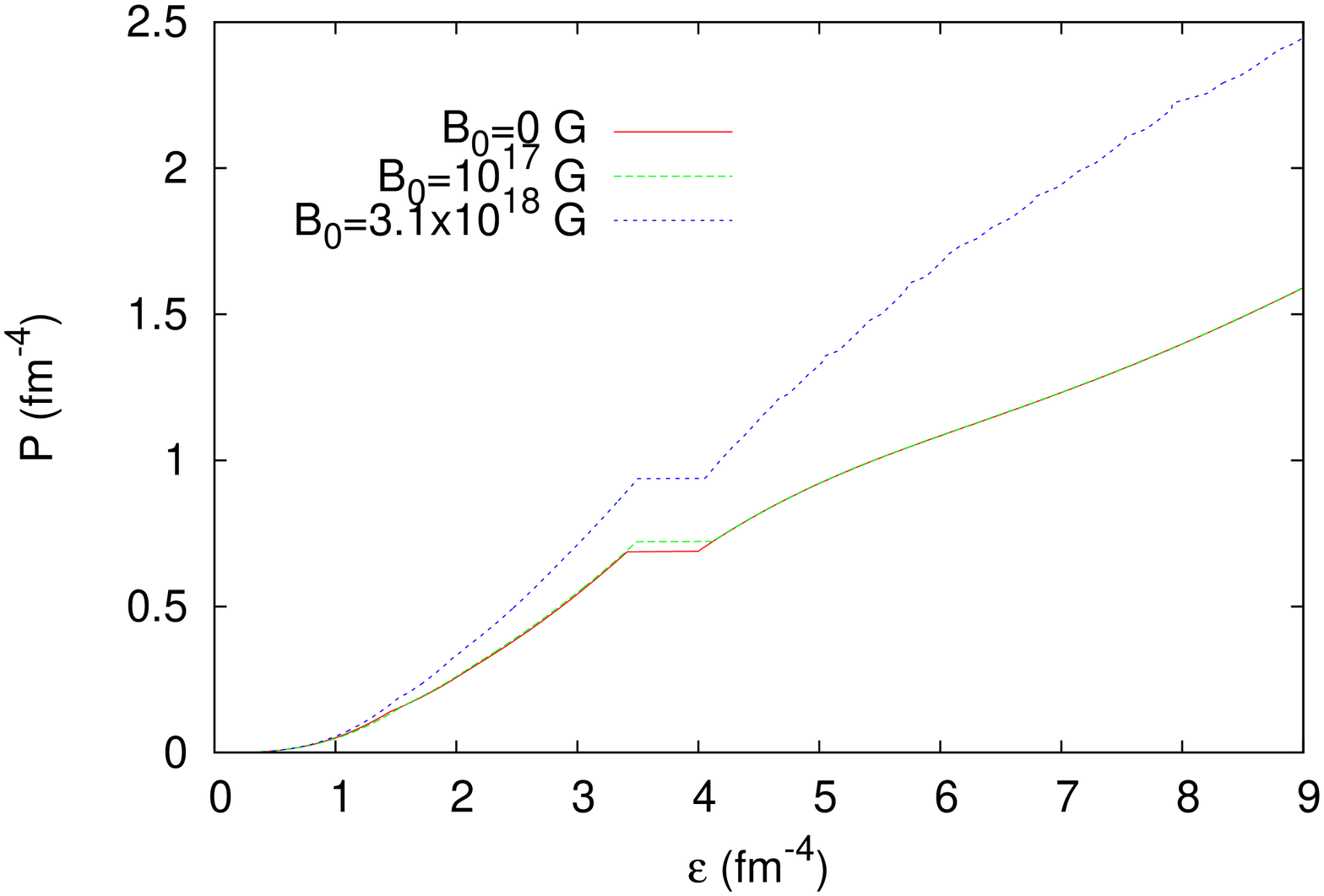}&
\includegraphics[angle=270,width=0.50\linewidth]{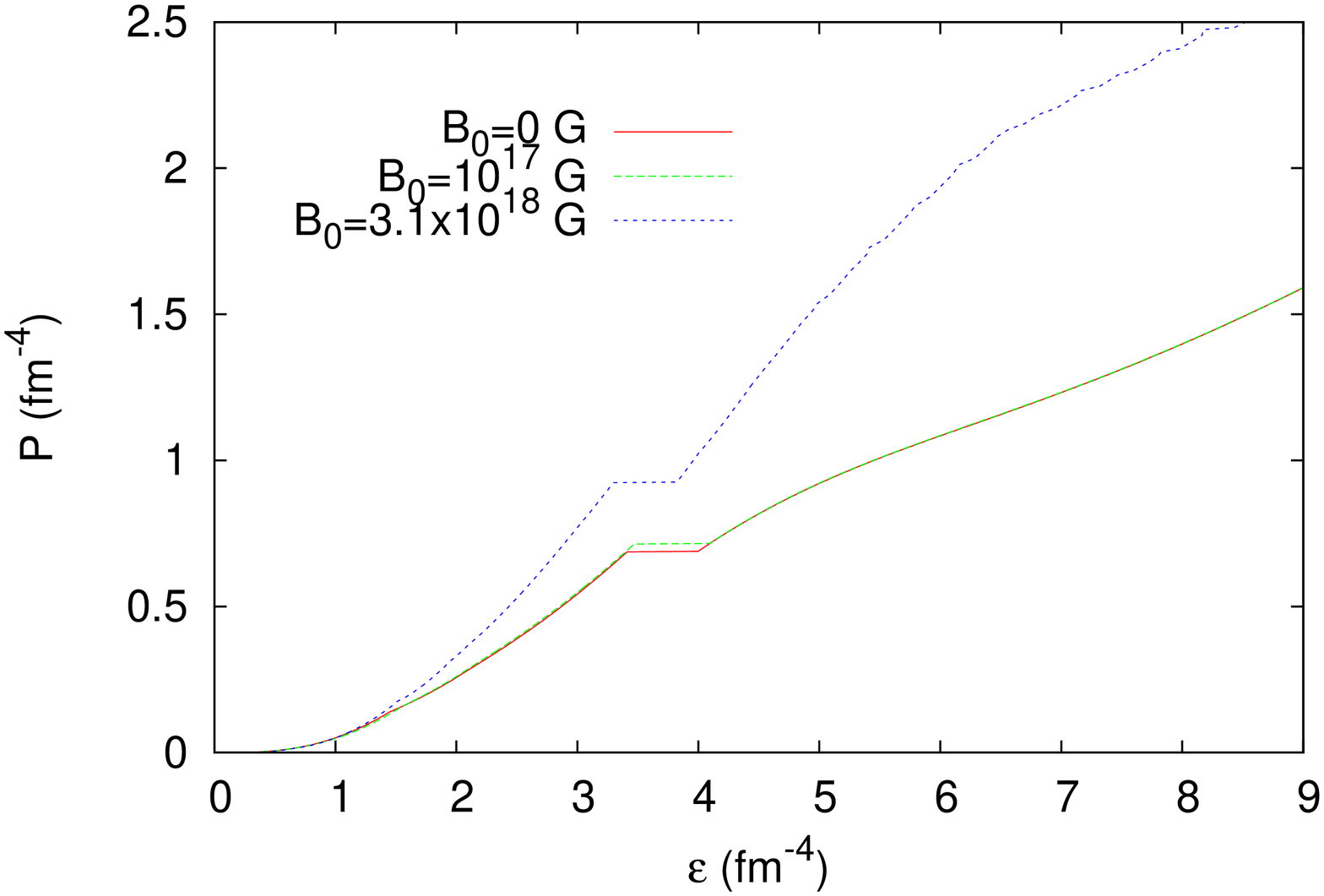} \\
\end{tabular}
\end{center}
\caption{EOS for the hybrid star without mixed phase built with the GM1 and SU(3) HK NJL parametrization, for slow (left panel) and fast (right panel) decays.}
\label{FIG-E}
\end{figure}
\begin{figure}
\begin{center}
\begin{tabular}{ll}
\includegraphics[angle=270,width=0.50\linewidth]{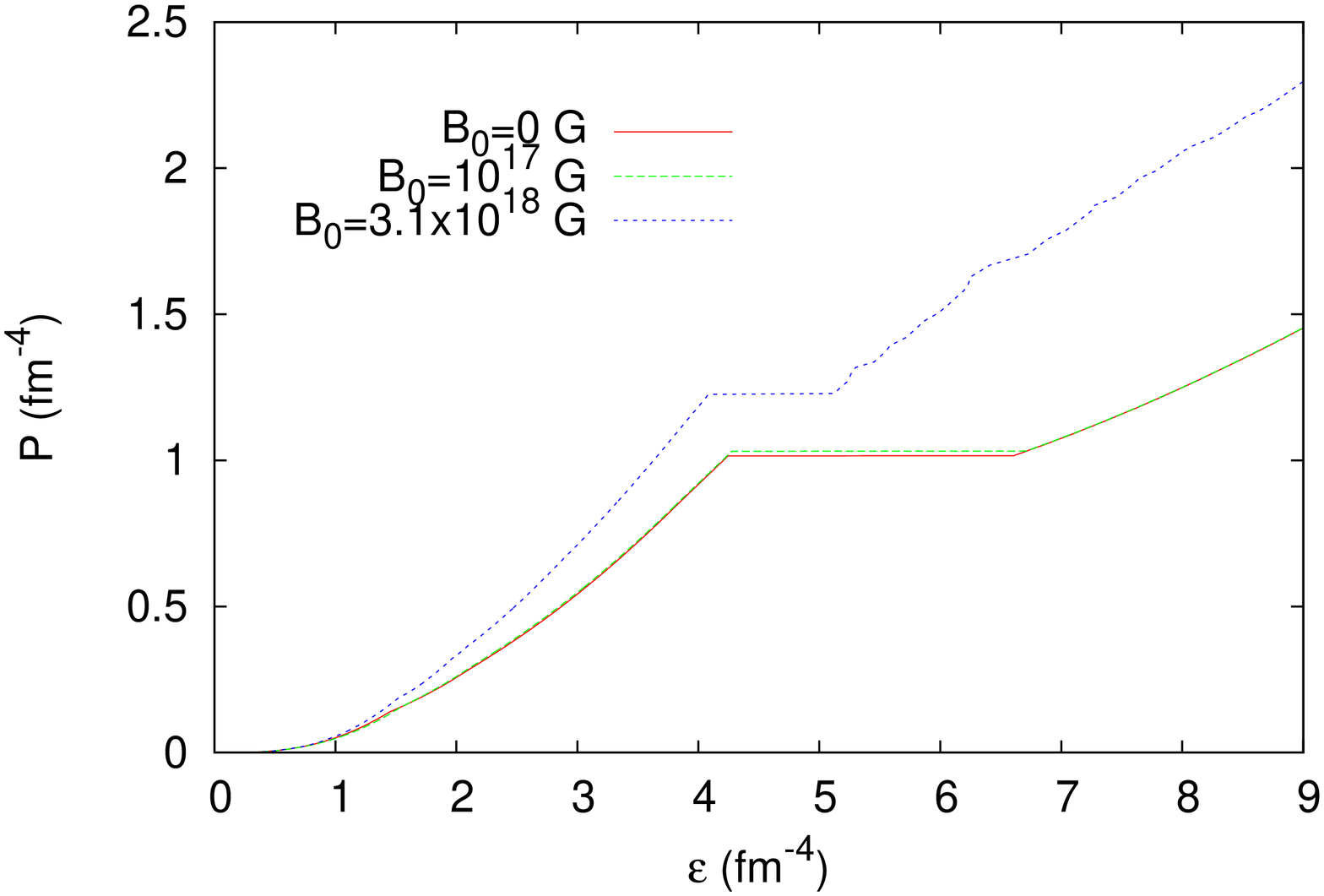}&
\includegraphics[angle=270,width=0.50\linewidth]{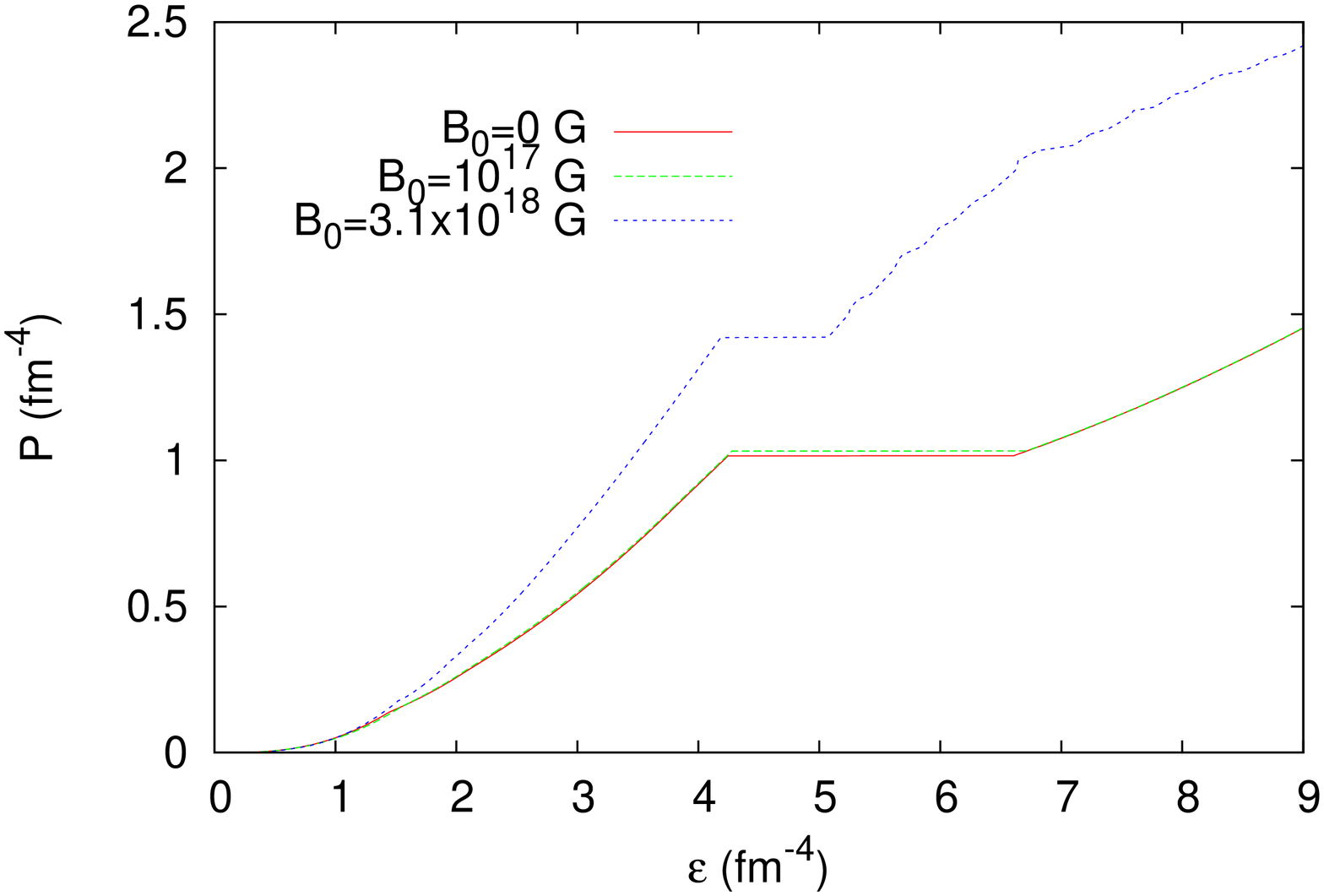} \\
\end{tabular}
\end{center}
\caption{EOS for the hybrid star without mixed phase built with the GM1 and SU(3) RKH NJL parametrizations, slow (left panel) and fast (right panel) decays.}
\label{FIG-Eb}
\end{figure}
\noindent In Figs.~\ref{FIG-E} and \ref{FIG-Eb}, we plot the EOS of hybrid stars under the influence of weak ($B_{0}=10^{17}~G$) and  strong ($B_{0}=3.1 \times 10^{18}~G$) magnetic fields. We note that as in the hadronic case the EOS for $B_{0}=10^{17}~G$ presents no great difference between FAST and SLOW cases. 
For the SU(3) HK (SU(3) RKH) parametrization, the onset of the quark phase occurs at $P\approx 0.69~fm^{-4}$, $\mu_{n}\approx1322$~MeV ($P\approx1.02~fm^{-4}$, $\mu_{n}\approx1419$~MeV) for $B_{0}=0~G$. For $B_{0}=10^{17}~G$, the onset of the quark phase (SU(3) HK) occurs at $P\approx 0.72~fm^{-4}$, $\mu_{n}\approx 1330$~MeV  regardless of the parametrization for FAST and SLOW cases. 
For the SU(3) RKH parametrization the onset of the quark phase occurs at $P\approx 1.031~fm^{-4}$, $\mu_{n}\approx 1422$~MeV for $B_{0}=10^{17}~G$ also for both FAST and SLOW cases. For $B_{0}=3.1\times 10^{18}~G$ the contribution of the magnetic field makes the EOS harder as compared with the EOS for $B_{0}=10^{17}~G$ and this effect is reflected in the higher values of the maximum masses. The presence of a strong magnetic field also affects the onset of the quark phase. In this case, the onset of the quark phase (SU(3) HK) occurs at $P\approx 0.92 fm^{-4}$, $\mu_{n}\approx 1261$~MeV and $P\approx 0.94 fm^{-4}$, $\mu_{n}\approx 1295$~MeV for FAST and SLOW cases, respectively. For the SU(3) RKH parametrization, the onset of the quark phase occurs at $P\approx 1.42 fm^{-4}$, $\mu_{n}\approx 1335$~MeV and $P\approx 1.22 fm^{-4}$, $\mu_{n}\approx 1352$~MeV for FAST and SLOW cases, respectively. From these results, we can determine in which way the magnetic field affects the values of pressure and chemical potential of the transition to the quark phase. We conclude that when the value of the magnetic field increases the pressure also increases, while the chemical potential decreases. This happens for FAST and SLOW cases. Also, one should notice that the transition to the quark phase, for not too low magnetic fields depend on the choice of the parametrization and on the way they vary inside the star, making the results very model dependent.

\begin{figure}[h]
\begin{center}
\begin{tabular}{ll}
\includegraphics[angle=270,width=0.50\linewidth]{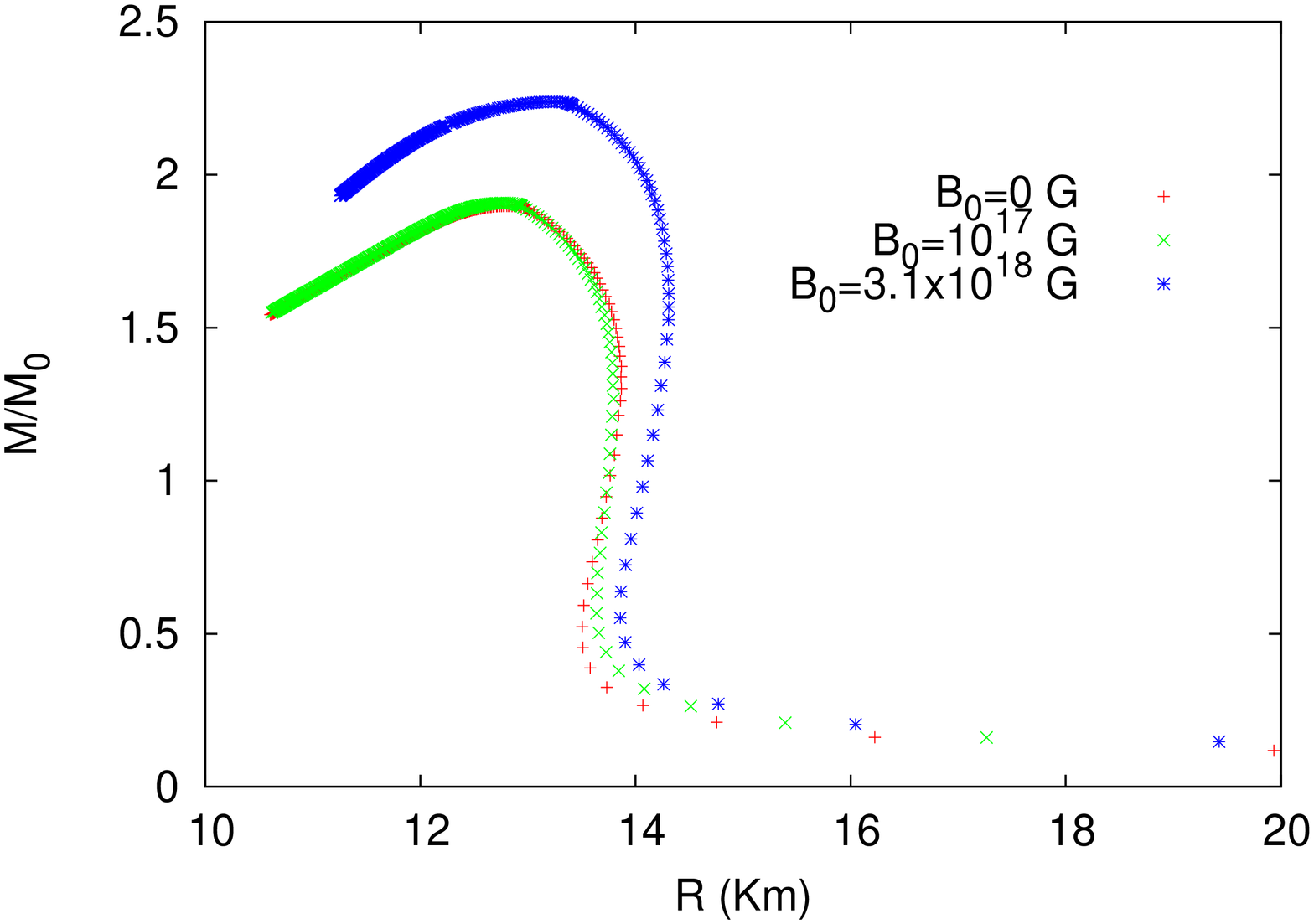}&
\includegraphics[angle=270,width=0.50\linewidth]{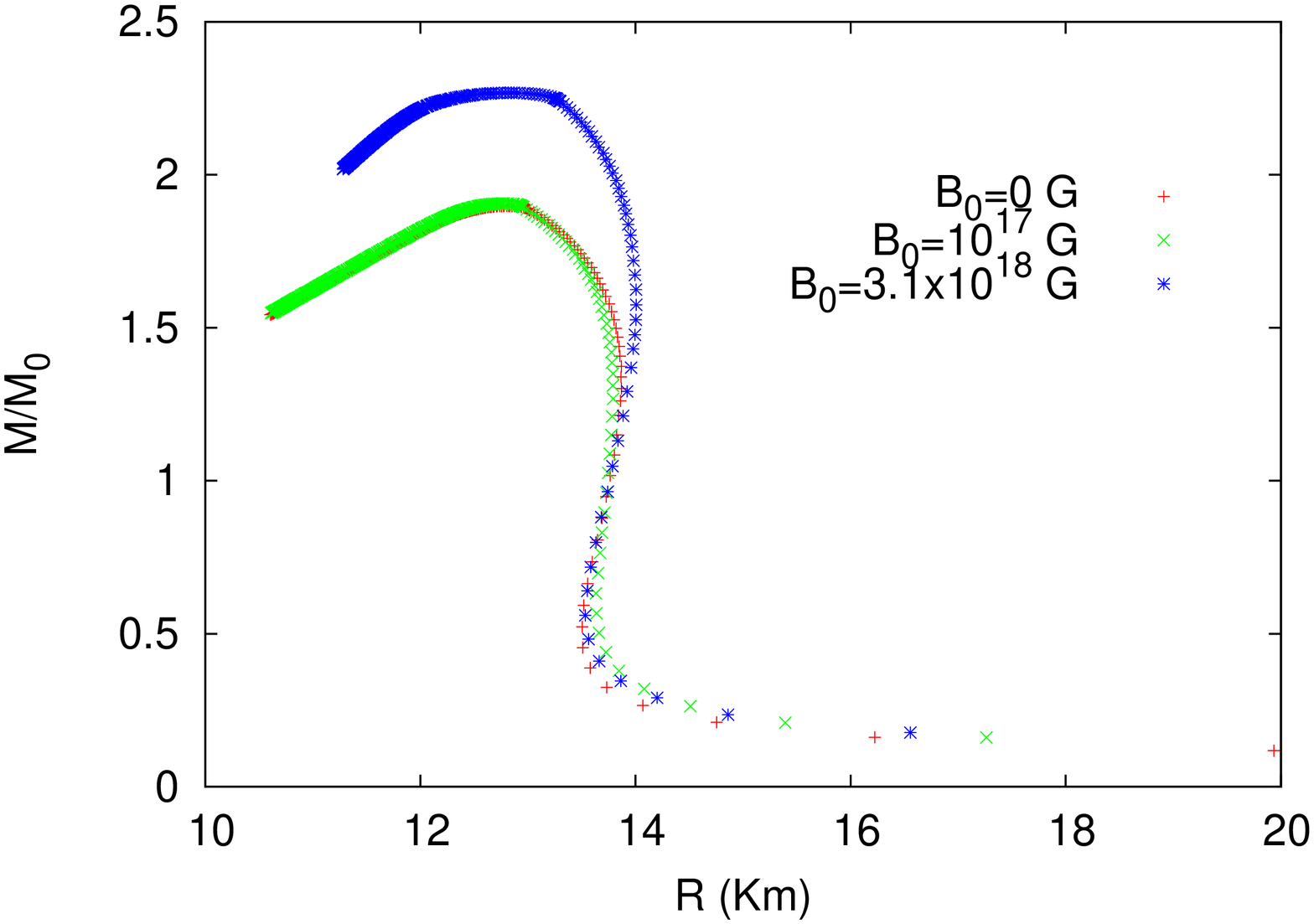} \\
\end{tabular}
\end{center}
\caption{Mass-radius curves for hybrid stars without mixed phase built with the GM1 and SU(3) HK NJL parametrizations, SLOW (left panel) and FAST (right panel) cases.}
\label{FIG-F}
\end{figure}

\begin{figure}[h]
\begin{center}
\begin{tabular}{ll}
\includegraphics[angle=270,width=0.50\linewidth]{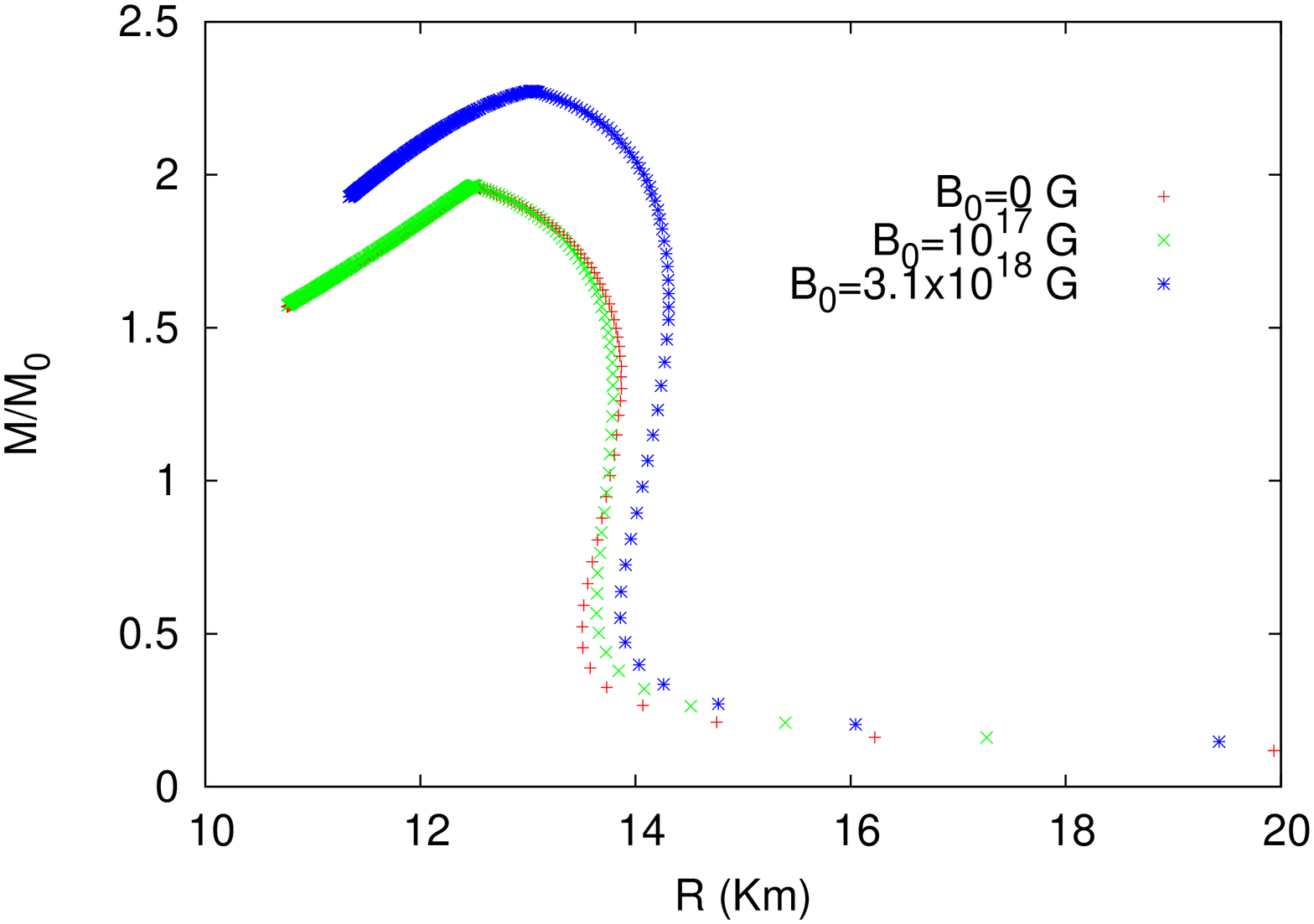}&
\includegraphics[angle=270,width=0.50\linewidth]{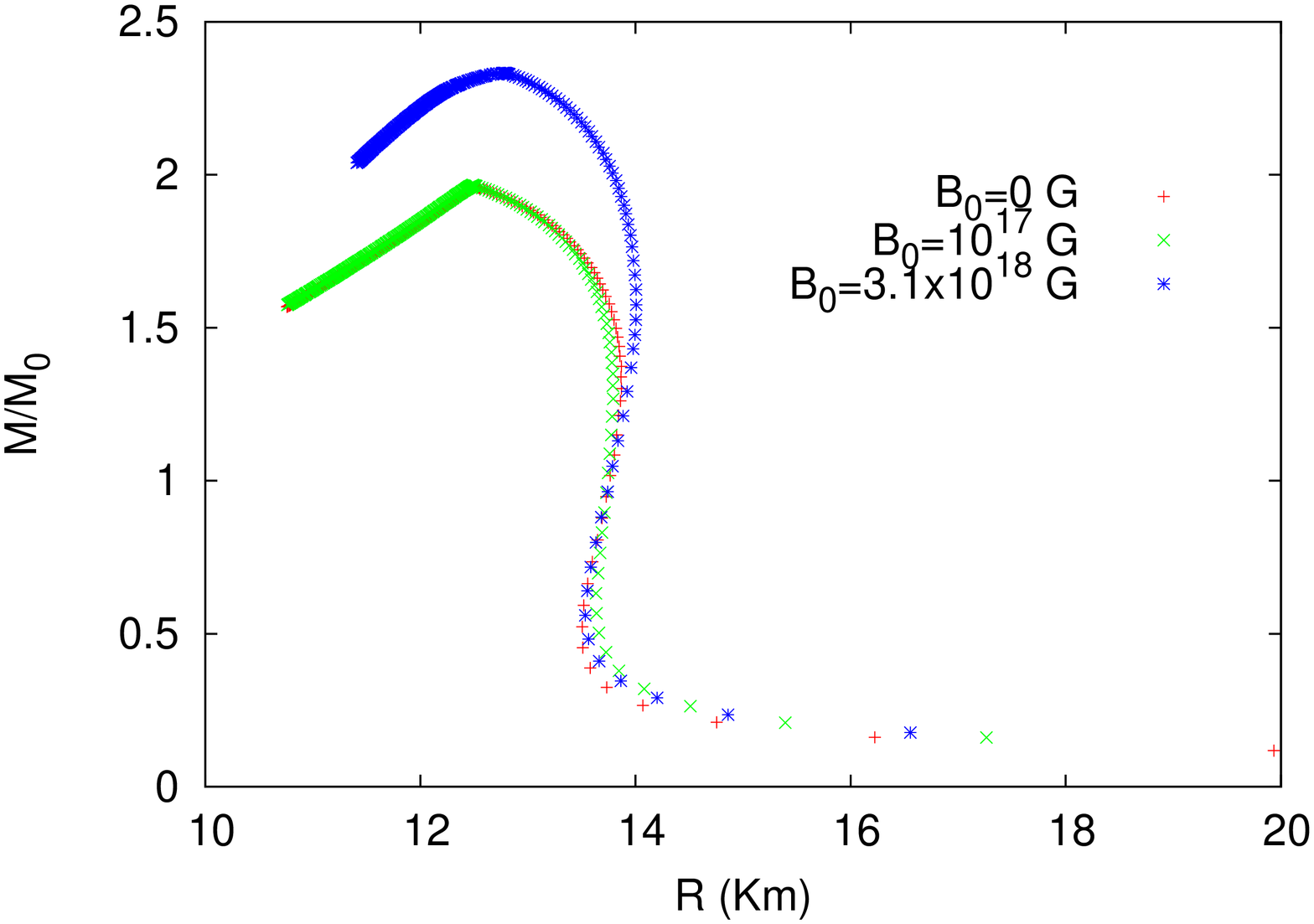} \\
\end{tabular}
\end{center}
\caption{Mass-radius curves for hybrid stars without mixed phase built with the GM1 and SU(3) RKH NJL parametrizations, SLOW (left panel) and FAST (right panel) cases.}
\label{FIG-Fb}
\end{figure}

In Fig.~\ref{FIG-F} and \ref{FIG-Fb}, we plot the mass-radius relation for both weak ($B_{0}=10^{17}~G$) and strong ($B_{0}=3.1 \times 10^{18}~G$) magnetic fields. As in the case of hadronic matter the tails of the hybrid stars were obtained with the insertion of the BPS EOS \cite{bps}. The values of the maximum masses and radii for a hybrid star are shown in TABLES \ref{table8} and \ref{table8b}. As expected, for $B_{0}=10^{17}~G$ the maximum masses and radii do not differ significantly from those found for $B_{0}=0$ in ref. \cite{marce}. It is worthwhile to mention that the small difference is consequence of considering a different central energy equal to $0.0001~fm^{-4}$ instead of $0.187 \times 10^{-9}~fm^{-4}$
as input to the TOV equations as done in \cite{marce}. From TABLES \ref{table8} and \ref{table8b} we can see that for $B_{0}=10^{17}~G$ the values of the maximum masses and radii for the two parametrizations of the magnetic field (FAST and SLOW cases) are the same. This result is also expected because the effects caused by the two parametrizations are not evident for a weak magnetic field, exactly as in the case previously discussed for hadronic stars.

Comparing the chemical neutron chemical potential $\mu_{n}$ of the onset of the quark phase with the one at the central energy density $\mu_{n}(\varepsilon_{c})$, we can check whether the star in hybrid or if it remains a hadronic star. We can see that for the SU(3) RKH parametrization only the case of $B_{0}=3.1\times10^{18}~G$ (SLOW case) is a hybrid star, all other case with this parametrization are hadronic stars (including the case of $B_{0}=0~G$), since $\mu_{n}>\mu_{n}(\varepsilon_{c})$. This does not happen when we use the SU(3) HK parametrization.  In this case all of the cases considered resulted in hybrid stars because 
$\mu_{n}<\mu_{n}(\varepsilon_{c})$.

The macroscopic properties of hybrids stars under effects of magnetic fields
have already been studied in the literature \cite{panda}. In ref. \cite{panda} the quark phase of the hybrid star is described by MIT bag model with $m_{u}=m_{d}=5.5$~MeV; $m_{s}=150$~MeV and two values for the Bag constant $(165~\mathrm{MeV})^{4}$ and $(180~\mathrm{MeV})^{4}$. Comparing our results with \cite{panda}, we obtain higher maximum masses and radii for weaker magnetic fields. For instance, in ref.~\cite{panda} for $B_{0}\approx2.2\times 10^{18}~G$ and $\mathrm{Bag}^{1/4}=180$~MeV the authors obtain $M=1.74~M_{0}$, $R=11.56$~Km (SLOW) and $M=1.72~M_{0}$, $R=11.49$~Km (FAST) and for $B_{0}\approx2.2\times 10^{18}~G$ and $\mathrm{Bag}^{1/4}=165$~MeV they obtain $M=1.72~M_{0}$, $R=9.80$~Km (SLOW) and $M=1.81~M_{0}$, $R=10.07$~Km (FAST). We obtain higher maximum masses and radii even for a weak magnetic field $B_{0}=10^{17}~G$. As it is well known, the mass and radius of compact stars obtained with the MIT bag model can be calibrated by increasing the value of the Bag constant. Nonetheless, it is important to emphasize that
it would be desirable that the values of the Bag constant were obtained through the study of stability windows \cite{jame2,jame} for which quark matter is
absolutely stable.
Therefore, even with strong magnetic fields, hybrid stars that contain a quark core described by the MIT model with Bag values
within the range of the stability windows, cannot support large maximum masses.
For instance, the stability windows for the MIT bag model are $147.0<\mathrm{Bag}^{1/4}<155.1$ and $152.1<\mathrm{Bag}^{1/4}<159.9$ for $B=0~G$ and $B=7.2\times10^{18}~G$, respectively \cite{jame2}. Note, however, that if different
corrections were included in the MIT model, as in \cite{sanjay},
hybrid stars with higher maximum masses could be attained.
Certainly,the stability conditions for quark matter should also be verified
when the quark phase is described by the NJL model. For zero temperature and
zero magnetic field, the NJL model is not absolutely stable either, but for
higher values of the magnetic fields, it is
consistent with the requirements for the existence of stable quark matter
\cite{jame,jame2}.


\begin{sidewaystable}
%\begin{table}[ht]
\begin{tabular}{|c|c|c|c|c|c|c|c|c|c|c|}
  \hline
Magnetic Field &\multicolumn{5}{ |c| }{FAST}&\multicolumn{5}{ |c| }{SLOW} \\
  \hline
&$M_{max}$ & R  & $\varepsilon_c$ &
$\mu_n(\varepsilon_{c})$ & R($M=1.4~M_{0}$) &
$M_{max}$ & R & $\varepsilon_c$ &
$\mu_n(\varepsilon_{c})$ & R($M=1.4~M_{0}$) \\ \cline{2-11}
 & ($M_{0}$) & (Km) & (fm$^{-4}$) & (MeV) & (Km) &
 ($M_{0}$) & (Km) & (fm$^{-4}$) & (MeV) & (Km) \\ \hline
$B=0$ G        & 1.90& 12.80 & 4.57 & 1360.8 & 13.86 &
1.90& 12.80 & 4.57 & 1360.8 & 13.86 \\ \hline
$B_{0}=10^{17}~G$& 1.91& 12.78& 4.57 & 1360.3 & 13.78 &
1.91& 12.77 & 4.62 & 1364.1 & 13.78\\ \hline
$B_{0}=3.1\times 10^{18}~G$ & 2.27& 12.82 & 4.69 & 1324.1 & 13.98 &
2.24 & 13.22 & 4.40 & 1324.1 & 14.29\\
\hline
\end{tabular}
\caption{ Maximum masses and related radii and central energy densities
for hybrid stars built with the GM1 and SU(3) HK NJL parametrizations. $\mu_{n}(\varepsilon_{c})$ is the chemical potential for neutron at the central energy density $\varepsilon_{c}$ and R($M=1.4~M_{0}$) is the radius for a $M=1.4~M_{0}$.
}\label{table8}
%\end{table}
\end{sidewaystable}

\begin{sidewaystable}
%\begin{table}[ht]
\begin{tabular}{|c|c|c|c|c|c|c|c|c|c|c|}
  \hline
Magnetic Field &\multicolumn{5}{ |c| }{FAST}&\multicolumn{5}{ |c| }{SLOW} \\
  \hline
&$M_{max}$ & R  & $\varepsilon_c$ &
$\mu_n(\varepsilon_{c})$ & R($M=1.4~M_{0}$) &
$M_{max}$ & R & $\varepsilon_c$ &
$\mu_n(\varepsilon_{c})$ & R($M=1.4~M_{0}$) \\ \cline{2-11}
 & ($M_{0}$) & (Km) & (fm$^{-4}$) & (MeV) & (Km) &
 ($M_{0}$) & (Km) & (fm$^{-4}$) & (MeV) & (Km) \\ \hline
$B=0$ G        & 1.96& 12.52 & 6.00 & 1402.7 & 13.86 &
1.96& 12.52 & 6.00 & 1402.7 & 13.86 \\ \hline
$B_{0}=10^{17}~G$& 1.97& 12.48& 4.29 & 1335.2 & 13.78 &
1.97& 12.49 & 4.29 & 1335.2 & 13.78\\ \hline
$B_{0}=3.1\times 10^{18}~G$ & 2.33& 12.79 & 4.69 & 1317.5 & 13.98 &
2.27 & 13.05 & 5.14 & 1354.5 & 14.29\\
\hline
\end{tabular}
\caption{ Maximum masses and related radii and central energy densities
for hybrid stars built with the GM1 and SU(3) RKH NJL parametrizations. $\mu_{n}(\varepsilon_{c})$ is the chemical potential for neutron at the central energy density $\varepsilon_{c}$ and R($M=1.4~M_{0}$) is the radius for a $M=1.4~M_{0}$.}\label{table8b}
%\end{table}
\end{sidewaystable}

\section{Conclusions}

In the present work we have revisited the calculations of magnetars composed of hadronic matter only as in \cite{bro,bro2,Stellar-matter-with} and also
composed of a quark core as in \cite{panda}. In the first case, the main targets were to compute the differences caused by the individual anomalous magnetic moments in the EOS with the inclusion of hyperons, their particle abundances and the resulting stellar properties. In the second case, our aim was to build a hybrid star with a quark core described by the NJL model, instead of the MIT bag model used in \cite{panda}.
We have chosen the Maxwell conditions to construct the hybrid star because our main concern was the evaluation of the macroscopic stellar
properties obtained from different models and it was already shown in
\cite{marce} that the Gibbs and Maxwell constructions result in practically the same results when the NJL model is used for the quark matter.

All calculations were performed with two different values for the magnetic field, $B=10^{17}$ G and $B=3.1 \times 10^{18}$ G, the last one being the stronger possible value, for which the anisotropic effects in the pressure can be circumvented, since we have opted to used an isotropic EOS.
The magnetic fields were chosen to be  density dependent and vary from a surface value of $B=10^{15}$ G to the two values mentioned above.

For the low value of the magnetic field, $B=10^{17}$ G , the results do not differ from the ones obtained for a non-magnetized star. These results are displayed in TABLES \ref{table4}, \ref{table8} and \ref{table8b} and were already expected. When
a strong magnetic field, $B=3.1 \times 10^{18}$ G, is
assumed, some conclusions can be drawn. For hadronic stars, the maximum masses increase with the inclusion of the anomalous magnetic moments, as expected, since
 they stiffen the EOS.
A fast decay mode for the density dependent magnetic field yields larger maximum masses, what is also seen in hybrid stars.
Generally, hybrid stars present lower maximum masses due to softer equations
of state, a well know result for non-magnetized stars and larger radii.
The central energy densities do not present a common pattern.
Moreover, with the models and constants chosen in the present work, we can describe the recently detected neutron stars with masses of the order of 2 $M_\odot$ \cite{demorest,antoniadis}, contrary to what was found, for instance in \cite{panda} with a quark core described by the MIT bag model and a not too large value of the magnetic field. We have also seen that a hybrid 
star cannot be always obtained with both parametrizations of the NJL model 
used in the present work for magnetized and non-magnetized matter. Although the EOS was built in such a way that the star could be hybrid, the TOV results have shown that the onset of the quark phases sometimes takes place
at energy densities higher than the ones found in the core of the star.

Finally, let's make some comments on the possible values of neutron stars radii.
Based on chiral effective theory, the authors of ref. \cite{Hebeler} estimate
the radii of the canonical $1.4M_\odot$ neutron star to lie in the range
9.7-13.9 Km. More recently, two different
analysis of five quiescent low-mass X-ray binaries in
globular clusters resulted in different ranges for neutron star radii.
The first one, in which it was assumed that all neutron stars have the same
radii, predicted that they should lie in the
range $R=9.1^{+1,3}_{-1.5}$ \cite{guillot}. The second calculation, based
on a Bayesian analysis, foresees radii of all neutron stars to lie in
between 10 and 13.1 Km~\cite{Lattimer2013}.
If one believes those are definite constraints, all hadronic
and hybrid stars with both zero and large magnetic fields
obtained with the choice of EOS studied in the present work
would be ruled out, as can be seen from
Figs. \ref{FIG-B} and \ref{FIG-F}. 
Nevertheless, as already
explained, the radii depend on the choice of the magnetic field decay rate.
Moreover, as pointed out in \cite{Lattimer2013}, better X-ray data is needed to
determine the compositions of accreting neutron
stars, as this can make 30\% or greater changes in inferred neutron star radii.

To conclude, let's say that as we have obtained our results for two limits
of the magnetic field, namely,
the lowest possible one that could contribute at least slightly to the EOS and
the maximum value that allows us to avoid anisotropic pressures, all possible
analysis for intermediate values are contemplated and the stellar maximum
masses will always lie in between the values we have calculated.
Based on previous experiences with the NJL model, we believe that our results
will not change qualitatively if another parametrization were used. However, had we chosen to introduce one of the possible vector interactions available in the literature \cite{njlv1} and \cite{njlv2}, the quark matter
EOS would certainly be harder and the consequences of including magnetic fields
are presently under investigation.

\begin{acknowledgments}
This work was partially supported by CAPES, CNPq and FAPESC (Brazil).
\end{acknowledgments}

\end{document}